\documentclass{emulateapj}

\usepackage{natbib}
\usepackage{epsfig}
\usepackage{amssymb}
\submitted{2007, ApJ, 671, in press}

\def\ifm#1{\relax\ifmmode#1\else$\mathsurround=0pt #1$\fi}

\def\msun{M_{\odot}}

\def\ltsima{$\; \buildrel < \over \sim \;$}
\def\lsim{\lower.5ex\hbox{\ltsima}}
\def\gtsima{$\; \buildrel > \over \sim \;$}
\def\gsim{\lower.5ex\hbox{\gtsima}}
\def\eg{{\it e.g.}\,}
\def\ie{{\it i.e.}\,}

\def\zs{$z_{\rm s}${ }}

\def\lensname{SDSSJ0737+3216{ }}
\def\zdvalue{0.3223{ }}
\def\zsvalue{0.5812{ }}

\def\re{$r_{\rm e}${ }}
\def\logre{$\log_{10} r_{\rm e}/{\rm kpc}${ }}
\def\stellarmass{$M^{*}${ }}
\def\logstellarmass{$\log_{10} M^{*}/\msun${ }}
\def\sindex{$n${ }}
\def\sersic{S\'ersic{ }}

\def\revalue{$0.59 \pm 0.007 {\rm \,stat} \pm 0.1 {\rm \,sys}\, {\rm kpc}${ }}
\def\stellarmassvalue{$2.0 \pm 1.0 {\rm \,stat} \pm 0.8 {\rm \,sys} \times 10^{9} \msun${ }}
\def\logrevalue{$-0.23 \pm 0.005 {\rm \,stat} \pm 0.07 {\rm \,sys}${ }}
\def\logstellarmassvalue{$9.3 \pm 0.2 {\rm \,stat} \pm 0.17 {\rm \,sys}${ }}

\def\bluesindexvalue{$1.0 \pm 0.1 {\rm \,stat} \pm 0.2 {\rm \,sys}$}

\def\SExtractor{{\sc SExtractor}{ }}

\def\nicmos{HST NICMOS{ }}
\def\nirc{NIRC2+LGSAO{ }}
\def\acs{HST ACS{ }}

\def\seeing{$\approx 0.4$ arcsec{ }}

\def\Bfilter{F555W{ }}
\def\Ifilter{F814W{ }}
\def\Hfilter{F160W{ }}
\def\Kfilter{K'{ }}

\def\Bband{\Bfilter-band{ }}
\def\Iband{\Ifilter-band{ }}
\def\Hband{\Hfilter-band{ }}
\def\Kband{\Kfilter-band{ }}

\def\magK{$m_{\rm Kp}${ }}

\def\psfzero{PSF0{ }}
\def\psfone{PSF1{ }}
\def\psftwo{PSF2{ }}
\def\psfthree{PSF3{ }}

\def\subschemeone{{\it sub}{ }}
\def\subschemetwo{{\it msub}{ }}

\def\pr{{\rm Pr}}
\def\pars{\mathbf{x}}

\def\data{\mathbf{d}}
\def\datai{d_i}
\def\datap{\mathbf{d}_{\rm p}(\pars)}
\def\datapi{d_{{\rm p},i}(\pars)}
\def\model{\mathsf{H}}

\shorttitle{Gravitational lens \lensname}
\shortauthors{Marshall et~al.}

\def\ucsb{1}
\def\lick{2}
\def\cfao{3}
\def\ut{4}
\def\harvard{5}
\def\mit{6}
\def\ucla{7}
\def\keck{8}
\def\kapteyn{9}
\def\jpl{10}
\def\victoria{11}

\begin{document}

\title{Super-resolving distant galaxies with gravitational telescopes: Keck-LGSAO and Hubble imaging of the lens system \lensname}

\author{Philip~J.~Marshall\altaffilmark{\ucsb}$^{*}$}
\author{Tommaso~Treu\altaffilmark{\ucsb}$^{\dag}$}
\author{Jason~Melbourne\altaffilmark{\lick,\cfao}}
\author{Rapha\"el~Gavazzi\altaffilmark{\ucsb}}
\author{Kevin~Bundy\altaffilmark{\ut}}
\author{S.~Mark~Ammons\altaffilmark{\lick,\cfao}}
\author{Adam~S.~Bolton\altaffilmark{\harvard}}
\author{Scott Burles\altaffilmark{\mit}}
\author{James~E.~Larkin\altaffilmark{\ucla}}
\author{David~Le~Mignant\altaffilmark{\keck,\cfao}}
\author{David~C.~Koo\altaffilmark{\lick}}
\author{L\'eon~V.E.~Koopmans\altaffilmark{\kapteyn}}
\author{Claire~E.~Max\altaffilmark{\lick,\cfao}}
\author{Leonidas~A.~Moustakas\altaffilmark{\jpl}}
\author{Eric~Steinbring\altaffilmark{\victoria}}
\author{Shelley~A.~Wright\altaffilmark{\ucla}}

\altaffiltext{\ucsb}{Physics department, University of California, Santa Barbara, CA 93106, USA} 
\altaffiltext{*}{{\tt pjm@physics.ucsb.edu}} 
\altaffiltext{$\dag$}{{Alfred P. Sloan Research Fellow}} 
\altaffiltext{\lick}{U.C.O./Lick Observatory, University of California, Santa Cruz, CA 95064, USA}
\altaffiltext{\cfao}{Center for Adaptive Optics, University of California, Santa Cruz, CA 95064, USA}
\altaffiltext{\ut}{Department of Astronomy \& Astrophysics, University of Toronto, Ontario, M5S 3H4, Canada; Reinhardt Fellow}
\altaffiltext{\harvard}{Harvard-Smithsonian Center for Astrophysics, Cambridge, MA 02138, USA}
\altaffiltext{\mit}{Department of Physics and Kavli Institute for Astrophysics and Space Research, Massachusetts Institute of Technology, 77 Massachusetts Ave., Cambridge, MA 02139, USA}
\altaffiltext{\ucla}{Division of Astronomy \& Astrophysics, University of California, Los Angeles, CA 90095, USA}
\altaffiltext{\keck}{W. M. Keck Observatory, Kamuela, HI 96743, USA}  
\altaffiltext{\kapteyn}{Kapteyn Institute, P.O. Box 800, 9700 AV Groningen, The Netherlands}
\altaffiltext{\jpl}{Jet Propulsion Laboratory, California Institute of Technology, 4800 Oak Grove Drive, MS-169/327, Pasadena, CA 91109, USA}
\altaffiltext{\victoria}{Herzberg Institute of Astrophysics, Victoria, British Columbia, V9E 2E7, Canada}


\begin{abstract}

We combine high-resolution images in four optical/infra-red bands,
obtained with the laser guide star adaptive optics system on the Keck
Telescope and with the Hubble Space Telescope, to study the
gravitational lens system \lensname (lens redshift~\zdvalue, source
redshift~\zsvalue). We show that (under favorable observing
conditions) ground-based images are comparable to those obtained with
HST in terms of precision in the determination of the parameters of
both the lens mass distribution and the background source. We also
quantify the systematic errors associated with both the incomplete
knowledge of the PSF, and the uncertain process of lens galaxy light
removal, and find that similar accuracy can be achieved with Keck
LGSAO as with HST. We then exploit this well-calibrated combination of
optical and gravitational telescopes to perform a multi-wavelength
study of the source galaxy at $0\farcs01$ effective 
resolution.

We find the \sersic index to be indicative of a disk-like object, but
the measured half-light radius (\re=\revalue) and stellar mass
(\stellarmass=\stellarmassvalue) place it more than three sigma away
from the local disk size-mass relation. The \lensname source has the
characteristics of the most compact faint blue galaxies studied, and
has comparable size and mass to dwarf early-type galaxies in the local
universe.
With the aid of gravitational telescopes to measure individual
objects' brightness profiles to 10\% accuracy, the study of the
high-redshift size-mass relation may be extended by an order of
magnitude or more beyond existing surveys at the low-mass end, thus
providing a new observational test of galaxy formation models.

\end{abstract}




\section{Introduction}
\label{sect:intro}

Galaxies do not appear in arbitrary combinations of luminosity, mass and
shape, but instead obey empirical scaling relations (such as the Fundamental
Plane for early-type galaxies). Explaining the origin, and cosmic
evolution, of the scaling relations is a fundamental goal of galaxy
formation theories.  

As far as disk galaxies are concerned, the hierarchical structure
formation scenario predicts a correlation between size and stellar
mass, with width depending on the distribution of the initial spin of
the dark halos \citep{F+E80}. At any given mass, the expected
distribution of sizes is well-approximated by a log-normal
distribution.  Qualitatively, this prediction is quite robust,
although the exact forms of the correlation and the distribution
depend on the details of baryonic processes such as energy feedback
from star formation and bulge instability
\citep{MMW98,She++03,Ton++06,Dut++07,S+B07}. Therefore, measuring the
shape and width of the correlation provides not only a test of the
standard paradigm, but also valuable information on the
poorly-understood baryonic processes happening at sub-galactic scales.

From an empirical point of view, the relation between size, luminosity
(or equivalently surface brightness) and stellar mass is well
established for disk galaxies in the local Universe
\citep[\eg][]{She++03,Dri++05}. Analysis of suitable objects in the 
Sloan Digital Sky
Survey shows that at any given mass (luminosity) the
distribution of galaxies is indeed well-approximated as log-normal, 
although the scaling with mass of the characteristic size
and the width of the distribution are non-trivial. Defining disk
galaxies as those being well-fit by a single \sersic component with index
$n<2.5$, \citet{She++03} find that above a characteristic stellar mass
($\log M_{*,0}/\msun\sim 10.6$ corresponding to approximately
M$_{r,0}=-20.5$), size scales rapidly with stellar mass ($R\sim
M_*^{0.39}$) and the scatter is relatively small ($\sigma_{\ln R}\sim
0.34$). Below the characteristic stellar mass the correlation flattens
($R\sim M_*^{0.14}$) and the scatter increases significantly
($\sigma_{\ln R}\sim 0.47$).

At intermediate redshift ($0.1\lsim z \lsim 1$) the nature and interpretation of the
size-luminosity or size-mass relation is more uncertain. Several
authors \citep[\eg][]{Fer++04,Bar++05,Tru++06,Mel++06} have used Hubble
Space Telescope images to determine the sizes of intermediate and high ($z \gsim
1$) redshift
galaxies, down to the resolution and completeness limits of HST
(roughly equivalent to 1~kpc and 10$^{10} \msun$). Recent studies
conclude, taking selection effects into account, that there is
significant evolution in the size-luminosity relation
\citep{Bar++05,Tru++06,Mel++06}. However, it is hard to disentangle
luminosity evolution from size evolution, to ensure that samples at
different redshifts are directly comparable, and to compare results
from different studies, as the selection criteria are often similar
but not identical (\eg color vs. morphology; morphology determined
via \sersic index vs.\ bulge to disk decomposition vs.\ concentration
parameter vs. visual classification). Overall, it appears that disk
galaxy evolution cannot be explained by pure luminosity or pure size
evolution, but requires a combination of both. In contrast, the
relation between size and stellar mass appears to have changed very
little since $z\sim1$ \citep{Bar++05}, much less than would be
expected in the naive model where stellar mass and size are
proportional to the virial mass and radius (and hence size is expected
to scale as $H(z)^{-\frac{2}{3}}$, where $H(z)$ is the Hubble
parameter). 
Rather, galaxies appear to be growing ``inside-out'' in scale radius 
as their stellar mass increases
such that the size-mass relation is preserved over cosmic time \citep{Bar++05}.
It has been suggested that galaxy evolution models that take into account the
ever-increasing concentration of
dark matter halos, and the further effect of baryons
via adiabatic contraction could provide the physics required to 
reproduce the observed trend
\citep{Som++06}, although this may make it more difficult to reproduce
simultaneously other scaling laws, for example the Tully-Fisher
\citep{T+F77} relation \citep{Dut++07}.

%
%
Lower mass ($M_* \lsim 10^{10} \msun$) galaxies are 
even less well understood. While the local 
size-mass relations of \citet{She++03} for low ($n<2.5$) and high ($n>2.5$)
\sersic index objects diverge, the interpretation of \sersic index as a
morphological galaxy classifier becomes more uncertain at lower masses 
\citep[e.g.\ ][]{CCD92,TBB04}. At the same time, 
the measurement of the structural
parameters themselves becomes harder as the galaxy size decreases.
Nevertheless, such small galaxies are important objects to understand: the
luminous compact blue galaxies first noted by \citet{K+K88} appear in large 
numbers at intermediate redshifts 
in deep HST images \citep[\eg][]{Noe++06,Raw++07}, but evolve very rapidly to
vanishing abundance in the local universe. What becomes of these objects, which
represent sites of small-scale but vigorous star formation, is a
topic of some debate, with dwarf spheroids \citep[\eg][]{Koo++94,Noe++06} and 
the bulges of disk galaxies \citep[\eg][]{Ham++01,Raw++07} the principle
candidates.

Gravitational lensing is a powerful tool with which
to extend the investigation of scaling
laws over cosmic time \citep[\eg][]{Tre07}. On the one hand, the lensing
geometry provides a precise and almost model-independent measure of total mass
of the lens galaxy. Since the lens galaxies are mostly early-type galaxies (or
the bulges of spirals), this gives a new handle on the mass profile of these
systems \citep{T+K04,Koo++06} and hence, for example, on the relationship
between stellar and total mass \citep{Bol++07}. On the other hand, the
background source is typically magnified by a factor of $\sim$10, mostly in the
form of a stretch along the azimuthal direction.  While lensing preserves
surface brightness, the increase in apparent size of the lensed source means
that the number of pixels at any one surface brightness also increases, such
that the isophotes are observed at higher signal-to-noise. Thus, gravitational
lenses act as natural telescopes, allowing one to gain a factor of $\sim10$ in
sensitivity and spatial resolution, and thus improve markedly our ability to
study the size and dynamical mass (through rotation curves) of intermediate and
high redshift galaxies. For example, studies of the internal structure of faint
blue galaxies~\citep{Ell97}, and in particular the most compact of these
\citep{Koo++94}, are currently limited by the resolution of HST \citep{Phi++97}.
When magnified by a gravitational lens, such objects become well-resolved. 
Thanks to the dedicated efforts of several groups, the
number of known gravitational lenses is increasing dramatically: it is now
possible to envision statistical studies of relatively large sample of lensing
or lensed galaxies in the near future.

In this paper we present multi-color high-resolution images of the
gravitational lens system \lensname \citep{Bol++06}, obtained with
both the Hubble Space Telescope and with the Laser Guide Star Adaptive
Optics (LGSAO) System on Keck II. The scientific goal of the analysis
of this case study is two-fold. First, we perform a detailed comparison
of the results of the lens modeling across bands, showing that -- when
a bright nearby star is available for tip-tilt correction and
conditions are favorable -- the most important parameters can be
measured with comparable accuracy with HST and Keck-LGSAO. Second, we
exploit this particular cosmic telescope to achieve super-resolution
of the source galaxy. \citep[See][ for Keck LGSAO observations of
a lens with a point-like source.]{McK++07} With a lens magnification of $\mu \gsim 10$, the
resolution of the HST and Keck images ($\sim 0\farcs1$ FWHM) 
corresponds to a physical scale of 
($0.66{\rm kpc} / \mu \approx 0.05{\rm kpc}$) at the
redshift of the source {\zs~=~\zsvalue}, comparable to the resolution
attainable from the ground when studying galaxies in the Virgo Cluster
in $1$~arcsec seeing. We derive the \sersic index, size, and stellar mass of
the source, and show that using gravitational telescopes the size-mass
relation may be extended by an order of magnitude in size with respect
to current studies, thus allowing one to probe, for example, whether
the change in slope and intrinsic scatter below the characteristic
mass persists to higher redshifts. 
  
This paper is organized as follows. After describing the observations
in section~\ref{sect:obs}, we outline in sections~\ref{sect:psf}
and~\ref{sect:subtraction} two sources of systematic error and our
strategies for dealing with them, before explaining our modeling
methodology in section~\ref{sect:modelmethod}. In
sections~\ref{sect:modelresults} and ~\ref{sect:source} we present our
results, which are then discussed (section~\ref{sect:discuss}) before
we draw conclusions in section~\ref{sect:conclude}. Throughout this
paper magnitudes are given in the AB system. 
We assume a concordance cosmology with matter and dark energy
density $\Omega_m=0.3$, $\Omega_{\Lambda}=0.7$, and Hubble constant
H$_0$=70 kms$^{-1}$Mpc$^{-1}$.


\section{Description of the observations}
\label{sect:obs}

\begin{figure*}[!ht]
\plottwo{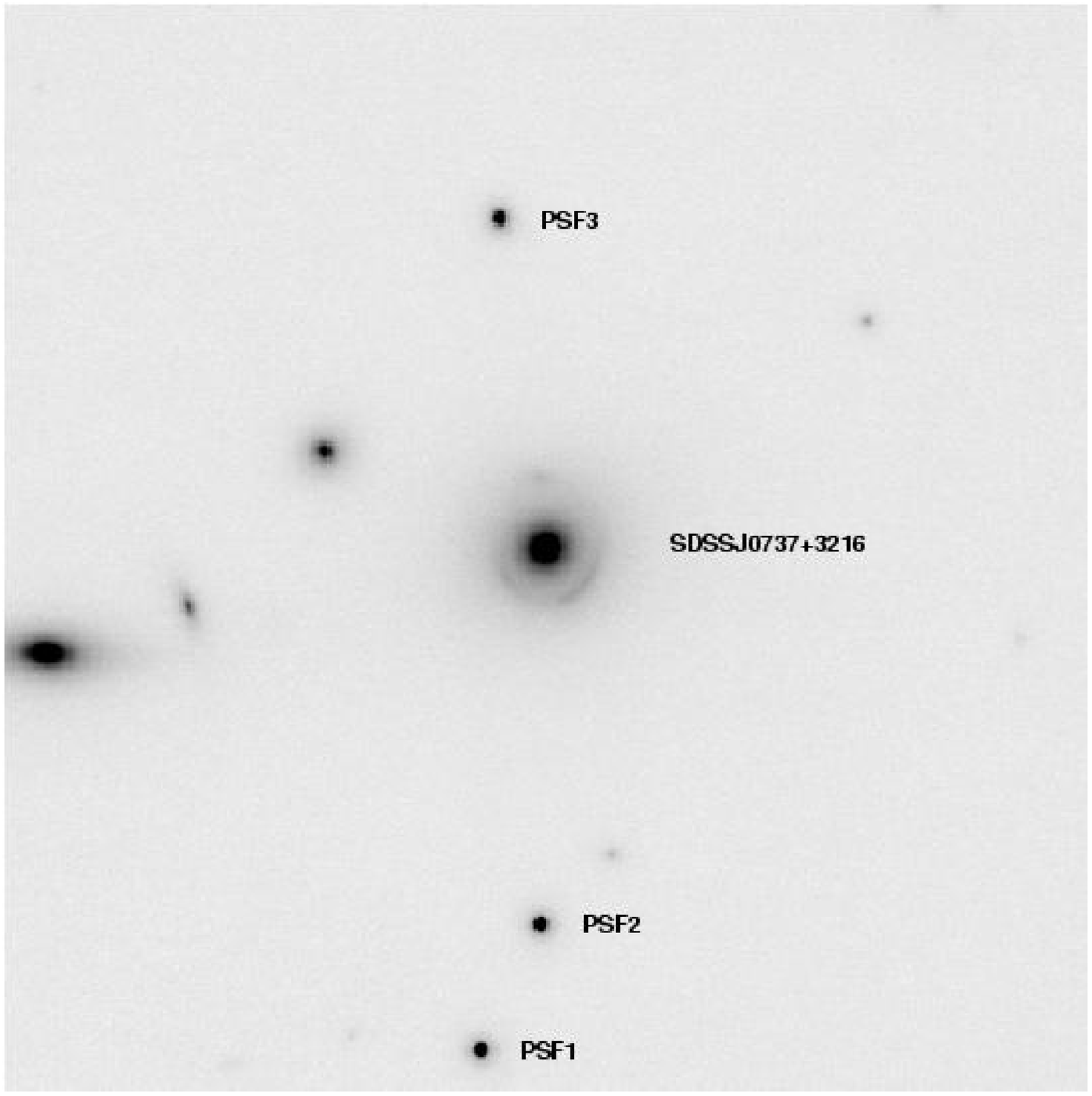}{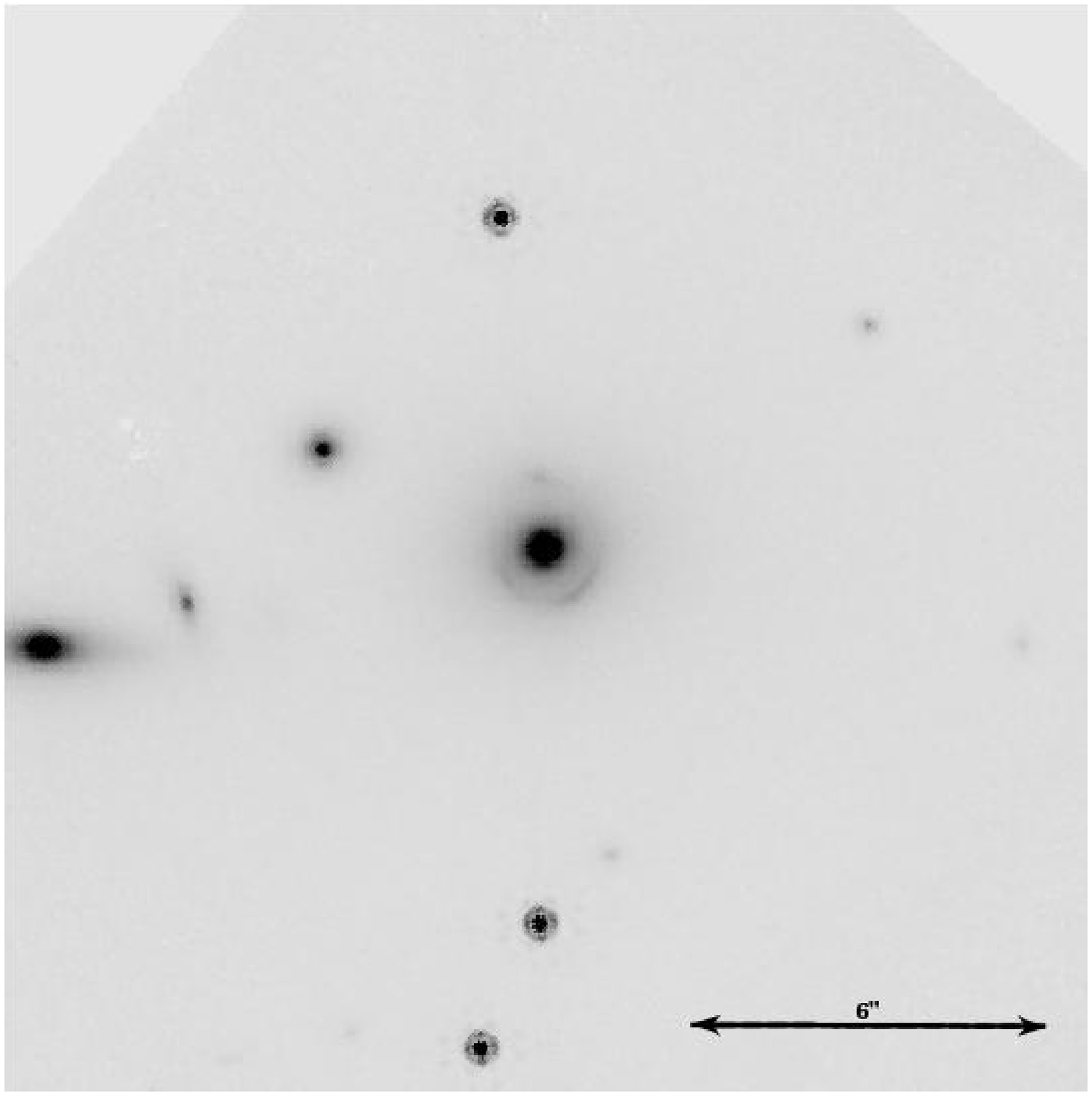}
\caption{\nirc \Kband (left) and \nicmos \Hband (right) images 
of \lensname, showing the stars used in the PSF modeling. A further PSF 
star was observed for the \nirc analysis.}
\label{fig:images}
\end{figure*}


\subsection{NIRC2 on Keck}
\label{sect:obs:nirc}

On December 11, 2006, we imaged \lensname with the LGSAO system on the
Keck II telescope.  The images were taken in the \Kband with the
near-infrared camera (NIRC2) in the wide field ($40\arcsec \times 40
\arcsec$) of view.  The pixel scale for this configuration is $0.04
\arcsec \; $pix$^{-1}$.  A total of 3120 seconds of exposure was
obtained.  To avoid saturating stars in the field, individual
exposures were 1 minute in duration (divided into two 30-second
co-adds).  A dither was executed after every set of 2 exposures to
improve the sky sampling.  Dithers were randomly chosen using a script
with a circular dither pattern of radius $3\arcsec$. The laser was
positioned at the center of each frame, rather than fixed on the
central galaxy.  Steinbring et al.\,(2007, submitted) demonstrate that
this method provides a more uniform AO correction over a larger area,
in comparison with the fixed laser method.  Observing conditions
during the run were good.

The observations were made as part of the Center for Adaptive Optics
Treasury Survey  (CATS, Larkin et al. 2007, in prep), which aims to image
$\sim 1000$ distant galaxies with Keck  adaptive optics.   The images were
processed with the CATS reduction procedure described in \citet{Mel++05}.  A
sky frame and a sky flat were created from the  individual science exposures
after masking objects.  Frames were then flat-fielded and sky-subtracted. 
The images were de-warped to correct for known camera distortions.  The
frames were aligned by centroiding on objects in the field, and finally
co-added to produce the final image.

The final processed image shows three unsaturated stars lying within  $10
\arcsec$ of the  lens position.  Two of these stars are between the tip-tilt
star and the lens.  The third lies on the opposite side of lens from the
tip-tilt star.   These stars provide a very strong  constraint on the
point-spread-function (PSF), which is often difficult to track for AO 
observations.  A further constraint on the PSF comes from observations of a
PSF  star-pair.  The star-pair observations were made immediately following
the lens observations. We picked a star pair that had a tip-tilt/PSF
orientation and separation similar to the  tip-tilt/lens system.  As a
result, the lens observation has  one of the best constrained PSFs ever
obtained for an extragalactic AO observation.  

A visual inspection of the stars in the field reveals an approximate double
Gaussian  profile, as used in simple models of adaptive optics PSFs
\citep[\eg][]{LSE06}.  The small-scale component of this profile was
observed to have a FWHM of  $\approx 2.5$ pixels, or $0.10$~arcsec.  This is
significantly larger than the diffraction limit of Keck in  the K-band
($\sim0.06\arcsec$), but is primarily the result of using the wide-field
camera which slightly under-samples the PSF.  The large-scale Gaussian
component has a FWHM of $\approx 0.40$~arcsec, indicating very good seeing.
From this simple PSF picture we estimate the Strehl ratio to be
approximately 0.35 for all the PSF stars, 
demonstrating consistently good AO
performance in these observations.


\subsection{ACS/NICMOS on HST}
\label{sect:obs:nicmos}

The lens system was observed with the Advanced Camera for Surveys
(ACS) and with the Near Infrared Camera and Multi Object Spectrograph
(NICMOS) on board HST on November 5 2006, as part of {\it HST} program
10494 (PI: Koopmans).  
One-orbit integrations were obtained through
filters F555W (2200s) and F814W (2272s) with the Wide Field Camera
centering the lens on the WFC1 aperture, \ie in the center of the
second CCD.
%
\footnote{Prior to the taking of these deep images,
shallow (420s) integrations were obtained with ACS
in both the F435W and F814W filters~\citep{Bol++06}, as part of the initial
SLACS snapshot program. These data are not 
used here due
to the low signal-to-noise and significant cosmic ray contamination, both of
which prevent detailed study of the faint ring.}
Four sub-exposures were obtained with a half-integer
pixel offset ({\tt acs-wfc-dither-box}) to ensure proper cosmic ray
removal and sampling of the point spread function. A one orbit
integration with the NIC2 camera through filter F160W was obtained
with NICMOS in {\tt multiaccum mode} for a total exposure time of
2560s. As with ACS, the integration was split in four
sub-exposures with a semi-integer pixel offset to ensure proper cosmic
ray/defect removal and improve sampling of the point spread function.

The ACS data were reduced using multidrizzle \citep{koe++02} as
described in \citet{Gav++07}. The NICMOS data were reduced using a set
of {\sc iraf} scripts based on the {\sc dither} package \citep{F+H02},
as described in \citet{T+K04}. The output pixel size was set to match
that of NIRC2 ($0\farcs0397$) to facilitate comparison between the HST
and reduced NIRC2 images.


\section{PSF characterization}
\label{sect:psf}

In order to predict accurately the data given a model lens image, we
must convolve it with the point spread function (PSF) of the
telescope. For the instruments on HST the PSF is calculable from the
engineering parameters that characterize the optics and detectors,
using the {\sc TinyTim} package \citep{K+H97}. However, the PSF varies
over time, both as a result of the ``breathing'' of the telescope over
the course of an orbit, but also monotonically as the system ages:
the Tiny Tim approximation is not always sufficient.

Somewhat similarly, the PSF derived from first principles for an
adaptive optics system is the sum of a Moffat profile for the seeing disk,
and the diffraction pattern due to the telescope itself. In practice,
the seeing, and the Strehl ratio, vary over the course of a set of
observations, making a priori predictions of the PSF of limited use.

In principle, one could include some variable parameters to describe
the model PSFs introduced above, and then fit for them simultaneously
with the lens model parameters. We show in
Section~\ref{sect:modelresults} that there is indeed enough
information in our data to constrain the PSF, thanks to the
multiple-images produced by the lens, but defer the 
investigation of model PSF parameters
to further work. Here we take a pragmatic approach and use nearby
unsaturated stars as estimates of the PSF at the position of the lens.
For the case of \lensname there are three suitable stars within
$\approx 10$~arcsec from the lens; we excised small cutout images of
these stars from the images from each instrument/filter
combination. The properties of these stars (henceforth referred to as
\psfone, \psftwo, and \psfthree) are given in
Table~\ref{tab:stars}. In addition, for the NIRC2 observations 
we used a
fourth star as described in section~\ref{sect:obs:nirc}. 
The use of any given one of these stellar model PSFs
constitutes an assumption which we can test using a statistical model
selection procedure we describe below.
                                                                                
\begin{table}[!ht]
\caption{Properties of stars used in the PSF characterization.
$\delta\theta_{\rm lens}$ is the angular separation between the PSF star and the
lens system / laser spot chip position.
$\delta\theta_{\rm TT}$ is the angular separation between the PSF star and the
tip-tilt star.
\psfzero was observed only during the AO run.}
\label{tab:stars}
\begin{center}
\begin{tabular}{cccccc}
\hline
ID             & RA      & Dec.        & $\delta\theta_{\rm lens}$ & $\delta\theta_{\rm TT}$ & \magK  \\
               & (J2000) & (J2000)     & (arcsec)                  & (arcsec)                  & (AB)  \\
\hline\hline
\psfzero  & 07:03:11.84  & -08:20:51.8 & 0.0                       & 17.8                     & 15.0 \\ 
\psfone   & 07:37:28.54  & +32:16:10.2 & 8.5                       & 14.7                     & 18.2 \\ 
\psftwo   & 07:37:28.46  & +32:16:12.3 & 6.3                       & 16.4                     & 18.1 \\ 
\psfthree & 07:37:28.51  & +32:16:24.2 & 5.6                       & 28.4                     & 18.3 \\ 
\hline
\end{tabular}
\end{center}
\end{table}

This phenomenological model has the advantage that it takes into
account the time-variability of the PSF as well as possible, providing
a simultaneous estimate of the PSF with the actual data. It also takes
into account the details of the image combination procedure in a
natural way -- whatever was done to the pixels of the lens image was
also done to the PSF. One disadvantage of our approach is the
introduction of additional noise -- however, the stars are
significantly brighter than the lens system and the pixel noise in the
PSF images can, we believe, be safely neglected. 
Three other disadvantages of our
approach are that the stellar spectra will not exactly match the
spectra of the lens or source galaxies within a given filter, nor will
the position of the PSF stars within a pixel exactly match the
intra-pixel centroiding of the lens or source galaxies, and nor will
the PSF at the position of the stars exactly match that at the lens
position. In the absence of a suitable interpolation scheme to solve
these problems, we resign ourselves to having just three models to
choose from, and attempt to infer the most appropriate one of the
three from the data. Following this procedure will give us an
indication of the relative importance of accurately knowing the
PSF. In other words, the variation of the results as a function of
adopted PSF will give us an indication of the systematic error
introduced by our approximate PSF. As we will show in the next
sections the parameters that we are interested in are fairly
insensitive to the choice of the PSF, and that our ignorance of
the the PSF is not the dominant source of error in our analysis.


\section{Lens galaxy subtraction}
\label{sect:subtraction}

As can be seen in Figure~\ref{fig:images}, the lens galaxy is much
brighter than the (lensed) source galaxy, and is a significant source of
contamination at the arc positions. The usual approach to this profile
is to subtract a smooth intensity distribution fitted to the lens
galaxy light. 
%
\citet{Bol++06} found it necessary to use a flexible B-spline model, combined
with careful manual masking of the multiple images, in order to obtain a
satisfactory removal of the lens light.
The problem
is that it is fundamentally very difficult to disentangle the light
coming from the lens galaxy from that coming from the source. 
\citet{Mou++06}
used the simpler elliptically-symmetric Moffat profile; a \sersic profile fit
could also have been performed.
To quantify this source of systematic uncertainty we investigate both
lens galaxy subtraction methods found in the literature, 
test them as best we can using the
data, and compare the results in terms of relevant lens and source
parameters.

In subtraction scheme \subschemeone we used a \SExtractor segmentation
map to mask out the detected pixels associated with the lensed images,
and then fitted an elliptically symmetric B-spline model with two
angular modes \citep[see the appendix of][ for details]{Bol++06}. In
this scheme, there is a danger that the tangentially-stretched images
will be truncated, leading to an overly-compact inferred source.  The
Moffat profile fit (henceforth referred to as subtraction scheme
\subschemetwo) was performed as in \citep{Mou++06}, with no masking of
the image. This model has the benefit of being somewhat more robust,
but must be expected to provide a much poorer quality of fit, leaving
more lens galaxy flux in the residual image and leading to a brighter,
larger inferred source.
%
Based on these considerations, we expect that
the two schemes will bracket the ideal solution and thus provide an
estimate of the systematic uncertainty.
A \sersic profile fit may well provide a better fit to the lens galaxy 
light than the Moffat profile: we use the Moffat profile in order to make our 
systematic error estimate a conservative one.


\section{Lens modeling methodology}
\label{sect:modelmethod}

Modeling of the images of extended sources lensed by galaxy-scale
lenses has been the subject of some considerable research in the last
few years \citep[see e.g.{}][]{W+D03,T+K04,D+W05,Koo05,Suy++06,B+L06,B+K07}. The
differences between these works revolve around the choice of
regularization scheme for the reconstructed source plane image, while
the lens models are largely consistent between the methods and reflect
the simplicity and consistency observed in gravitational lens
potentials~\citep{Koo++06}. The regularization is important due to the
very large numbers of parameters employed to describe the source plane
intensity.

In this work, and in a previous article~\citep{Mou++06}, we choose to
model the source galaxy using simply-parametrized
elliptically-symmetric \sersic profile components. We pursue this
approach for two reasons. Firstly, images of intermediate and high
redshift galaxies very often show morphologies representable by
collections of simply-parameterized components (bulges, disks,
star-forming regions etc.). The second reason is that we seek a
quantitative understanding of galaxy luminosity, mass, size and shape
as a function of redshift, and this is best achieved by analyzing the
image data within the context of a sensible phenomenological model
(the \sersic profile). The resulting inferences will of course be
model-dependent (by design), and we should expect the corresponding
precision to be high as a result of the additional information used in
the fit. Most importantly, our results will be directly comparable to
other photometric and morphological studies. After all, a pixel based
reconstruction will have to be fit by a parameterized \sersic model in
order to derive shape and luminosity parameters that can be compared
with the literature.

%
For our lens models we follow previous authors
and use the singular isothermal ellipsoid
(SIE) model~\citep[e.g.,][]{KSB94}. 
A number of authors \citep[e.g.,][]{T+K04,R+K05,Koo++06}  
have shown the SIE model to provide a very good approximation of
the lens potential on galaxy scales. 
%
%
The basic lens equations describing the deflection
of light by this model can be found readily 
elsewhere \citet[\eg][]{KSB94,E+W98,E+W01,Sch06} and are
not repeated here. Suffice to say that given the deflection angle as a
function of lens plane position, the corresponding source plane
position can be rapidly calculated, using the formulae in
\citet[\eg,][]{E+W01}.  The price we pay for this high computation
speed is a significant systematic error in the source parameters as
inferred through the lens. The intrinsic spread of logarithmic density
slopes (where the SIE profile has slope $m=1$) is approximately 0.12, based on
the large sample of strong lenses analysed by 
\citet{Koo++06}; in the appendix we show that this gives rise to a
fractional uncertainty in source size of about 12\%, and an error in
the inferred source magnitude of 0.26. Implementing a more flexible
lens model would translate this systematic error into a statistical
one -- while this is beyond the scope of this paper we note that a
reasonable goal is to reduce all other systematic errors to below the
level set by the lens mass profile.

Since our source surface brightness distribution is the analytic
\sersic profile, we can compute the source intensity at each desired
source plane position, and assign it to the original image plane pixel
value -- we do this on a twice sub-sampled grid to reduce rounding
errors.  \citep[This simple but effective ``poor man's ray-tracing''
is described further in][]{SEF92}.  In this way a predicted image can
be calculated for any given set of lens and source parameters. Before
comparison with the data image we convolve the model image with a PSF
image (derived from the image of a nearby star, as described in
Section~\ref{sect:psf} above). With the PSF image grid being much
smaller than the data image grid the speed of the computation is
greatly increased.

The $N$-pixel model image~$\datap$ and data image~$\data$ are compared
via the likelihood function:
\begin{eqnarray}
\pr(\data|\pars) &= \frac{1}{Z_{\rm L}} \exp{\left( -\frac{\chi^2}{2}\right)},\label{eq:lhood} \\
{\rm where}\; \chi^2 &= \sum_i^N \frac{(\datapi - \datai)^2}{\sigma_i^2}, \\
{\rm and}\; Z_{\rm L} &= (2\pi)^{N/2} \prod_i^N \sigma_i.
\end{eqnarray}
This form contains an implicit assumption of uncorrelated Gaussian
pixel noise, which is well-justified for the background-limited Keck
data. When using the HST images, we note that the counts are always
such that the Gaussian approximation to the Poisson distribution is
always valid, and compute the uncertainties $\sigma_i$ from the square
root of the image itself. We account for the correlated noise
introduced by the drizzling procedure by computing the 
{\it equivalent single pixel noise} \citep{Cas++00}, essentially by 
reducing the
uncertainties by a factor close to the fourth power of the ratio
between the output and input pixel scales. This has the effect of
making the reduced chi-squared approximately equal to unity in the case of a
good fit. In principle one could estimate the pixel covariance matrix and use
that in the calculation of $\chi^2$, at greater computational expense. We leave
this to future work, and note that the correlated errors are unlikely to affect
our statistical error bars by more than a factor of two. As we shall see,
systematic errors are of greater concern.

Our simple lens model has 5 parameters: position ($x$ and~$y$), velocity
dispersion~$\sigma_{\rm SIE}$,\footnote{While the strong lensing image
separation is a direct measurement of the mass enclosed by the Einstein radius,
when working with the SIE model the overall normalisation is more conveniently
described by the single parameter~$\sigma_{\rm SIE}$. This has the added
benefit of being (more or less) straightforwardly connected 
to dynamical mass estimates from spectroscopic velocity 
dispersions~\citep[\eg][]{T+K02}}
mass distribution ellipticity (defined as $\epsilon =
(1-q^2)/(1+q^2)$ where $q$~is the axis ratio), and orientation angle. 
We assign
uniform prior PDFs on the latter three; for the lens centroid we take the center
of the lens light as our best guess, and assert a Gaussian prior PDF  of  width
one pixel centered on this position. Similarly, for the source position we assign
a Gaussian prior PDF of width 0.1~arcsec centered on the lens position. 
(Since we know that the source is 
lensed, and into a almost circular ring at that, we know that the source
position must be very close to the optical axis. The Gaussian prior does allow
for putative source positions at larger radii, but has the effect of 
sensibly down-weighting those
models which are unlikely to provide a good fit. The value of $0.1$ comes purely
from experience with looking at lens models and simulated lenses.)
However, we assign uninformative uniform priors for the orientation~$phi$, 
\sersic index~$n$,
effective radius~$\theta_{\rm e}$, and source magnitude (where the logarithmic
nature of this quantity captures our even greater prior ignorance). For the
ellipticity we assume the standard weak lensing intrinsic ellipticity
distribution, a Rayleigh distribution of mean 0.25. 
%
(Note that the relation between the effective radius~$\theta_{\rm e}$, 
effective semi-major
axis~$a_{\rm e}$ and axis ratio~$q_{\rm s}$ is 
$\theta_{\rm e} = a_{\rm e} \sqrt{q_{\rm s}}$, so that our effective radii may be
compared directly with the ``circularised'' radii of \eg
\citeauthor{She++03}~\citeyear{She++03}).
We shall see in
sections~\ref{sect:lensresults} and~\ref{sect:sourceresults} that our choices
of prior PDF have very little influence on the posterior inferences. These are
defined by the joint posterior PDF:
\begin{equation}
\pr(\pars|\data,\model) =
\frac{\pr(\data|\pars,\model)\pr(\pars|\model)}{\pr(\data|\model)}.
\label{eq:posterior}
\end{equation}
$\pr(\pars|\model)$ is the product of the individual prior PDFs referred to
above. We sample the unnormalized numerator of equation~\ref{eq:posterior} using
the multi-purpose Markov Chain Monte-Carlo code {\tt BayeSys} \citep{Ski04}, a
robust package used in a number of other cosmology and lensing analyses
\citep[\eg][ and Jullo et al 2007, in prep.]{Odm++04,Mar06,Lim++06}.  

The symbol $\model$ in equation~\ref{eq:posterior} represents the set
of assumptions that go into the inference of the parameters via the
MCMC analysis. Such models can be compared quantitatively using the
evidence, $\pr(\data|\model)$. This statistic is calculated during the
initial ``burn-in" period of the sampler, and, while dominated by the goodness
of fit, does take into account the different prior PDFs that might be
employed. For further reading about evidence analysis we recommend
\citet{MacKay}.

In this work, the prior PDFs are kept fixed while different PSF models
and lens galaxy subtraction schemes are tried, an approach also followed by
Suyu et al (2007, in prep.). A simple ranking could
be achieved by using some different monotonic function of the
chi-squared statistic; we note here though that the correct weights to
use when combining parameter estimates from different analyses are
exactly the evidence values (provided all models are deemed equally
probable {\it a priori}). This can be seen by marginalizing the
parameter posterior PDF over the models -- each individual model's
posterior gets multiplied by its (renormalized) evidence during the
summation:
\begin{equation}
\pr(\pars|\data) = \sum_{\model} \pr(\pars|\data,\model)\hat{\pr}(\data|\model).
\label{eq:modelaveraging}
\end{equation}
In practice, one model often has much higher evidence than the others on offer,
meaning that the sum can be approximated by this single term: this is model
selection. 


\section{Lens modeling results}
\label{sect:modelresults}

\begin{figure*}[!p]
\vspace{-4\baselineskip}
\begin{center}\begin{minipage}{0.8\linewidth}
  \plotone{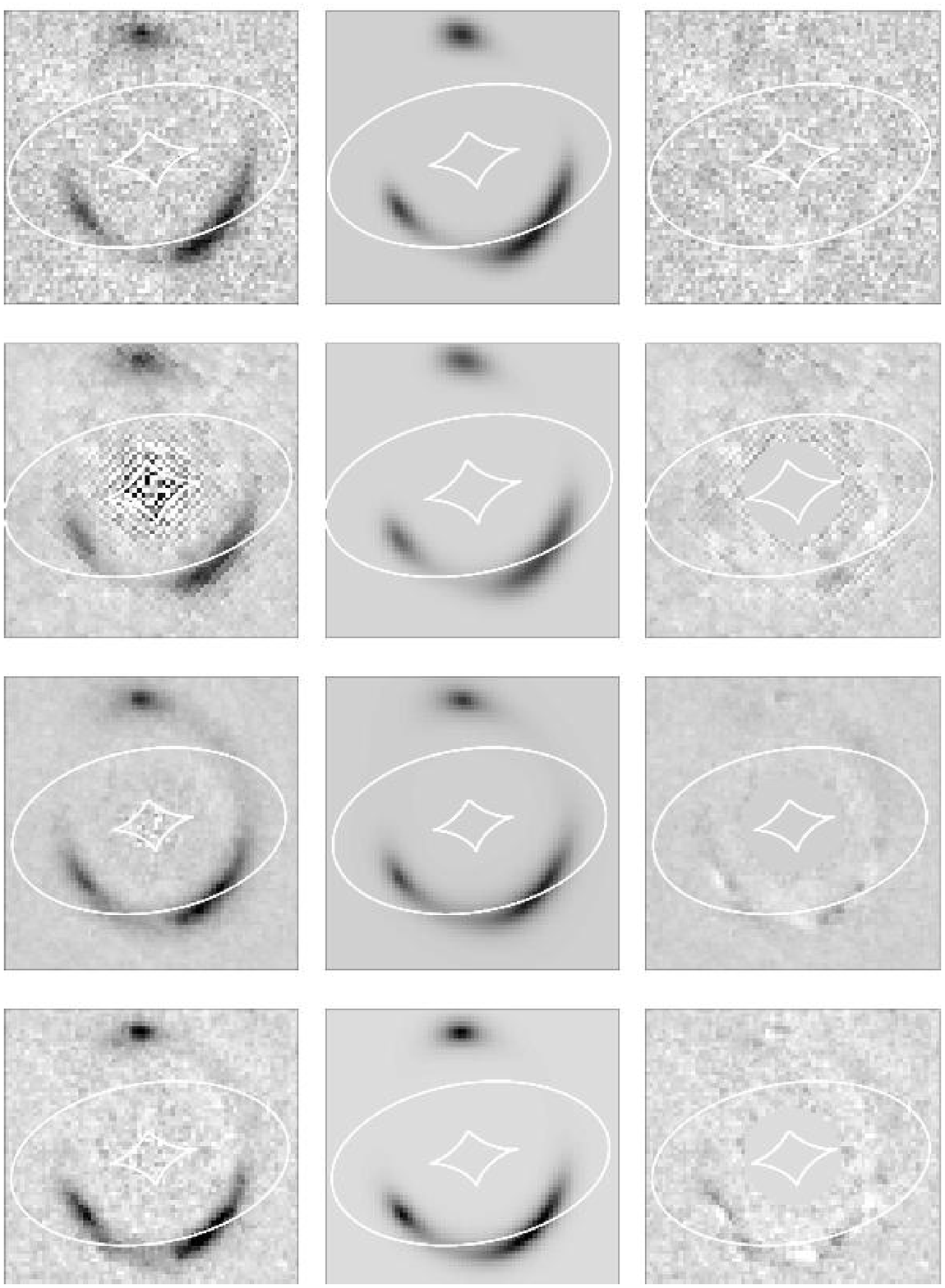}
\end{minipage}\end{center}
\vspace{2\baselineskip}
\caption{Data (left panels), predicted data (middle panels) 
and residual (right panels) for 
the best-fit lens models. Top row: \nirc \Kband data; second row: \nicmos data;
third row: \acs \Ifilter data; bottom row: \acs \Bfilter data.
The critical curve and asteroid caustic of the lens model are
overlaid in each case. 
The optimal PSF model was used for each dataset, and the lens galaxy subtraction
scheme was \subschemeone.
The pixel scale is 0.0397 arcsec: all these
cutout images are 2.81 arcsec on a side.}
\label{fig:fits}
\end{figure*}

Figure~\ref{fig:fits} shows the fits to the four imaging datasets introduced
above. For subtraction scheme \subschemeone (see section~\ref{sect:subtraction},
shown in the Figure) 
the residuals are close to zero, with little significant structure 
in the residual
images (especially in the infra-red filters). 
We show the results of the
statistical model selection analysis in Table~\ref{tab:stats}, for all datasets. 

We find that the different PSF models are easily differentiated (top half of the
table), with typical evidence ratios of a few tens. This is reflected in the
chi-squared statistic,which is not surprising given that the parameter space
volumes are identical between the different PSF models. The relative evidence is
determined almost entirely by the goodness of fit, which is significantly 
better for \psfzero in the case of the NIRC2 data. 
%
\footnote{We put the reduced chi-squared values in context by computing the
number of sigma, $N_{\sigma}$, by which the unreduced chi-squared~$\chi^2$ 
deviates from the mean of its distribution. We do
this using Fisher's approximation, that $\sqrt{2\chi^2}$ is 
Gaussian-distributed with mean~$\sqrt{2 k - 1}$ and unit variance, where~$k$ is
the number of degrees of freedom, assumed to be large.}
This may be due to the shape of the PSF at the lens being better matched by a
stellar image at the same position relative to the laser spot (which PSF0
provides).
For the HST datasets, the 
most appropriate PSF star to use varies between filters, as we might expect.

The situation with the lens galaxy subtraction schemes is less clear: here the
goodness of fit is dominated by the lens galaxy model such that we cannot use
the evidence straightforwardly to select the most appropriate model {\it for the
source galaxy}. The limiting case would be a lens galaxy model so flexible that
all the flux was subtracted, leaving a zero-flux inferred source and a
chi-squared of zero. What we can take from Table~\ref{tab:stats} is that the low
goodness of fit associated with subtraction scheme \subschemetwo indicates that
a significant amount of lens galaxy flux is being
left un-subtracted, a conclusion vindicated by inspection of the 
residual images (not shown).
The different schemes provide us with a rough estimate of the contribution of
lens galaxy subtraction to our systematic error budget. 

A side effect of the domination of the lens galaxy subtraction problem is that
the reduced chi-squared values from the lens modeling are often not  close to
unity. However, this need not affect our conclusions about the PSF model for fixed
subtraction scheme: a good PSF is required at all 4 image positions, but the
galaxy subtraction residuals vary between these points.

\begin{table}[!t]
\caption{Model selection statistics for each analysis.}
\label{tab:stats}
\begin{tabular}{cccccccc}
\hline
Dataset   & Subtraction     & PSF       & $\hat{\chi}^2$ & $N_\sigma$ & Relative   \\
          & scheme          & model     &                &            & evidence  \\
\hline\hline
\nirc     & \subschemeone   & \psfzero  & 1.219 & 10.6 & 1.0 \\
\Kfilter  & \subschemeone   & \psfone   & 1.220 & 10.7 & 0.03 \\
          & \subschemeone   & \psftwo   & 1.222 & 10.8 & 0.001 \\
\hline
\nicmos   & \subschemeone   & \psfone   & 0.991 & -0.46 &   1.0   \\
\Hfilter  & \subschemeone   & \psftwo   & 0.990 & -0.52 & 193     \\
          & \subschemeone   & \psfthree & 0.991 & -0.43 &   0.2 \\
\hline
\acs      & \subschemeone   & \psfone   & 6.276 & 153.3  &  1.0 \\
\Ifilter  & \subschemeone   & \psftwo   & 6.259 & 153.0  &  $e^{44}$  \\
          & \subschemeone   & \psfthree & 6.277 & 153.4  &  0.4  \\
\hline
\acs      & \subschemeone   & \psfone   & 1.083 & 4.16  &  1.0 \\
\Bfilter  & \subschemeone   & \psftwo   & 1.084 & 4.18  &  0.07 \\
          & \subschemeone   & \psfthree & 1.084 & 4.20  &  0.01 \\
\hline\hline

\nirc     & \subschemeone   & \psfzero  & 1.219 & 10.6 & 1.0 \\
\Kfilter  & \subschemetwo   & \psfzero  & 2.704 & 65.7 & $e^{-3900}$ \\
\hline
\nicmos   & \subschemeone   & \psftwo   & 0.989  & -0.52 & 1.0 \\
\Hfilter  & \subschemetwo   & \psftwo   & 2.596  & 62.3  & $e^{-4200}$ \\
\hline
\acs      & \subschemeone   & \psftwo   & 6.259  & 153 &  1.0  \\
\Ifilter  & \subschemetwo   & \psftwo   & 297  & 2650 &  $e^{-23500}$  \\
\hline
\acs      & \subschemeone   & \psfone   & 1.083  & 4.2 &  1.0  \\
\Bfilter  & \subschemetwo   & \psfone   & 1.666  & 29.6 &  $e^{-1500}$ \\
\hline\hline
\end{tabular}
\vspace{\baselineskip}
\end{table}

\begin{figure*}[!p]
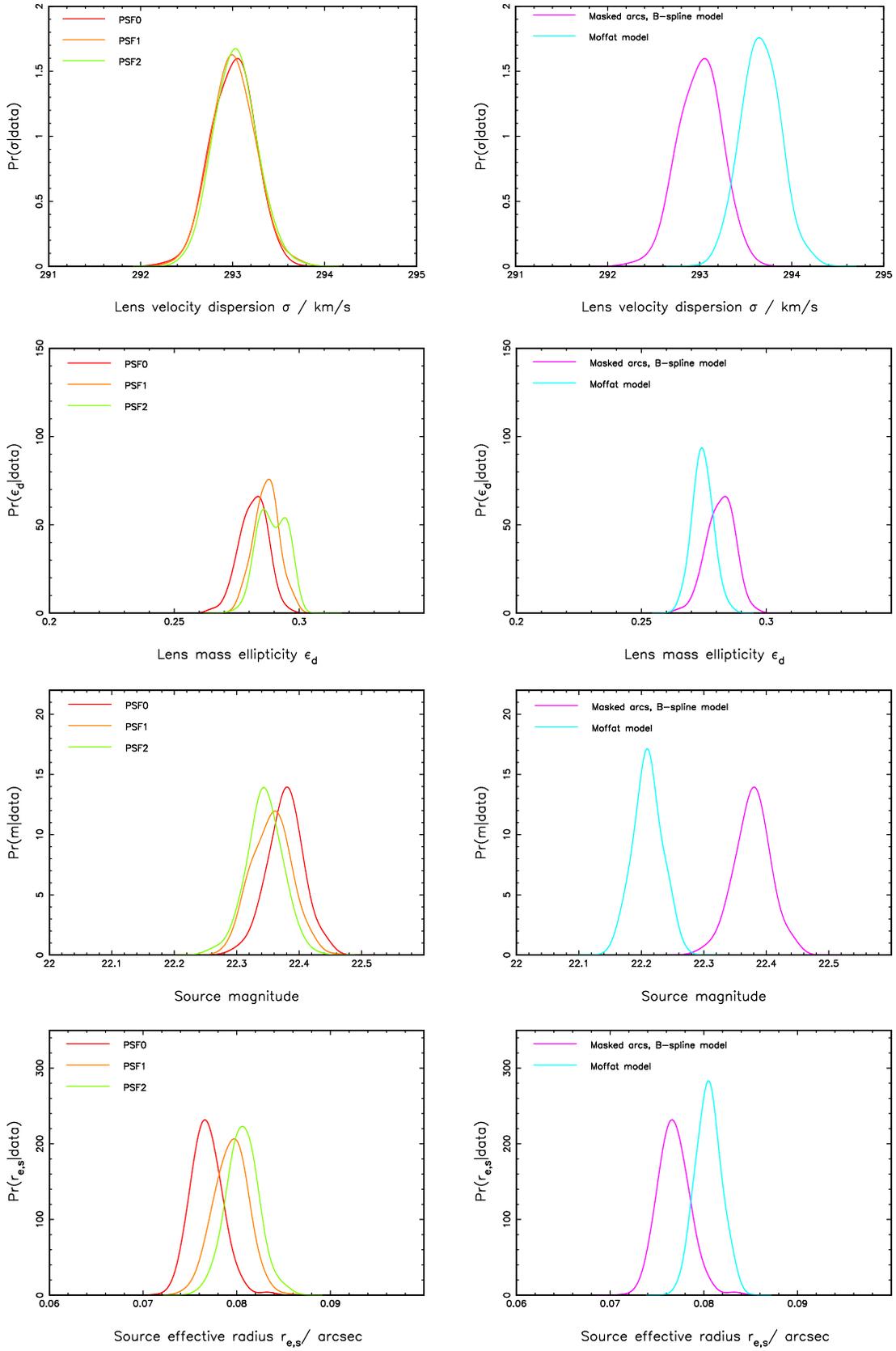
  
\begin{center}\begin{minipage}{0.9\linewidth}
  \plottwo{lens_sub_psf_comparison_vdisp.lineplot.ps}{lens_psf0_subtraction_comparison_vdisp.lineplot.ps}
\end{minipage}\end{center}
\begin{center}\begin{minipage}{0.9\linewidth}
  \plottwo{lens_sub_psf_comparison_eps.lineplot.ps}{lens_psf0_subtraction_comparison_eps.lineplot.ps}
\end{minipage}\end{center}
\begin{center}\begin{minipage}{0.9\linewidth}
  \plottwo{lens_sub_psf_comparison_b_mag.lineplot.ps}{lens_psf0_subtraction_comparison_b_mag.lineplot.ps}
\end{minipage}\end{center}
\begin{center}\begin{minipage}{0.9\linewidth}
  \plottwo{lens_sub_psf_comparison_b_re.lineplot.ps}{lens_psf0_subtraction_comparison_b_re.lineplot.ps}
\end{minipage}\end{center}
\vspace{\baselineskip}
\caption{Marginalized posterior probability 
%
distributions for four of the model parameters, given the \nirc data only. 
Top row: lens SIE velocity dispersion.
Second row: lens mass ellipticity.
Third row: source AB magnitude. 
Bottom row: source effective radius.
Left panels: comparing different PSF models. Right panels: 
comparing different lens galaxy subtraction schemes (\subschemeone dark,
\subschemetwo light). }
\label{fig:systematicpars}
\end{figure*}

Figure~\ref{fig:systematicpars} shows the 1-d marginalized probability distributions
for a selection of lens and source model parameters, 
given the \nirc infra-red imaging dataset, in order to illustrate the effect of
the different PSF models and the different lens galaxy subtraction schemes on
the inferences. Similar results were obtained from the other filters' data, and
are not shown here for the sake of clarity.

This figure shows that the choice of PSF model is not critical in
determining the available accuracy on the model parameters: in all cases the
parameter estimates agree within the statistical precision. The choice of lens 
galaxy subtraction scheme has a more significant
effect on the model parameters; in particular, the two schemes investigated give
rise to a difference of $\sim0.2$~magnitudes in source brightness.

To marginalize over the range of PSF models  one would use the relative
evidence values to weight the different posterior PDFs (as shown in
equation~\ref{eq:modelaveraging}); however, since the
evidence ratios in the top half of Table~\ref{tab:stats} are typically
significantly different from unity we approximate this procedure by simply
selecting the PSF model with the highest evidence.  For the rest of this
paper, we use the optimal PSF models for each dataset (from the maximum
evidence values given in Table~\ref{tab:stats}), and assert the \SExtractor
detected object mask subtraction scheme \subschemeone: the alternative
\subschemetwo distributions given in  Figure~\ref{fig:lenspars} (and those for
the other model parameters) provide estimates of the systematic errors we
expect for each parameter. We now compare parameter estimates in the four
different filters to compute the properties of the lens and the source.


\subsection{Lens properties}
\label{sect:lensresults}

\begin{figure}[!ht]  
\plotone{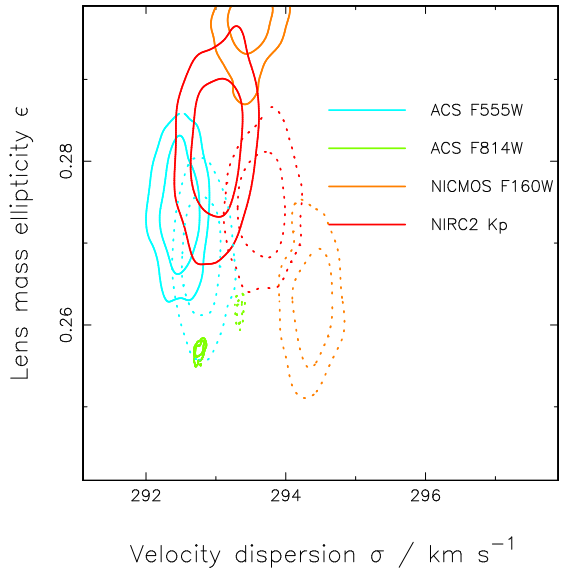}
\caption{
%
Marginalized joint posterior probability distributions, given each dataset, 
for the lens SIE velocity dispersion and
mass distribution ellipticity. 
The contours enclose 68\% and 95\% of the integrated probability.
Solid curves are for
the preferred galaxy subtraction scheme \subschemeone, while the dashed curves
are for the alternative scheme \subschemetwo.}
\label{fig:lenspars}
\end{figure}

Figure~\ref{fig:lenspars} shows the inferred SIE velocity dispersion and mass
distribution ellipticity for the \lensname lens. These parameters (along with
the mass orientation, not shown) agree reasonably well across the filters, as
they should given the  achromaticity of the lensing effect. The largest
discrepancies come from the deeper \acs \Ifilter image. The likelihood
function for this data is steeper, making it both harder for an MCMC sampler
to explore the parameter space, and for a simple model to provide a good fit. In this case the
inferred parameter uncertainties should be accepted with caution. 
Still, the inferred SIE velocity dispersion is in good agreement with that found by
\citet{Koo++06} from their shallower HST/ACS snapshot data.

We note that an offset of $0.5$~km/s in the velocity dispersion is equivalent
to one of 3.4~milliarcsec in the Einstein radius, a fractional error of
$0.3\%$. We assume that the reported image platescales are known to better than
this, but this may not be the case.  The truncation of the posterior pdf for
lens ellipticity is a direct result of our assumption of a prior on this
parameter that was uniform between 0.0 and 0.3. The lack of strong degeneracy
between ellipticity and any other parameter indicates that this truncation is
not a problem in this case -- but it serves as a warning for future analyses.



\section{Source properties}
\label{sect:source}

\begin{figure*}[!ht]  
\plotone{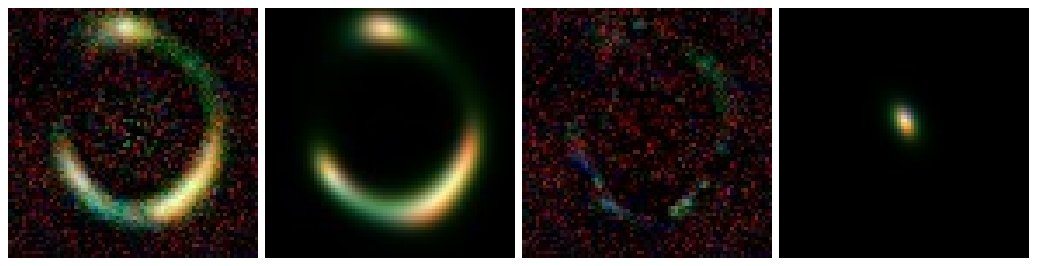}
\caption{Multi-filter reconstruction of the source behind \lensname. 
From left to right we plot: data, predicted data, residual and reconstructed
source plane images for 
the best-fit lens models, assuming optimal PSF model and lens galaxy subtraction
scheme~\subschemeone. Note the resolution of a red, compact core centered on
a more
extended blue light distribution. The red, green and blue image channels are
given by the \Kband, \Iband and \Bband images respectively, and the relative
scales were chosen (manually) to equilibrate the noise levels across the channels. }
\label{fig:colorjpgs}
\end{figure*}

Having calibrated the optics of our cosmic telescope we turn our
attention to the target of the observation: the lensed source at redshift~\zs.
Figure~\ref{fig:colorjpgs} shows the multi-color reconstruction of this object, 
which shows the presence of a red, compact core centered on
a more
extended blue light distribution. The ellipticity and orientation of the source
are a good match with those found from shallower data by \citet{Koo++06}. 
We note that the alignment of the different filters' reconstructions is very
good, and that qualitatively we seem to be recovering the large-scale stellar
component rather than being dominated by any smaller-scale features.


\subsection{Source photometry and morphology results}
\label{sect:sourceresults}

\begin{figure*}[!ht]
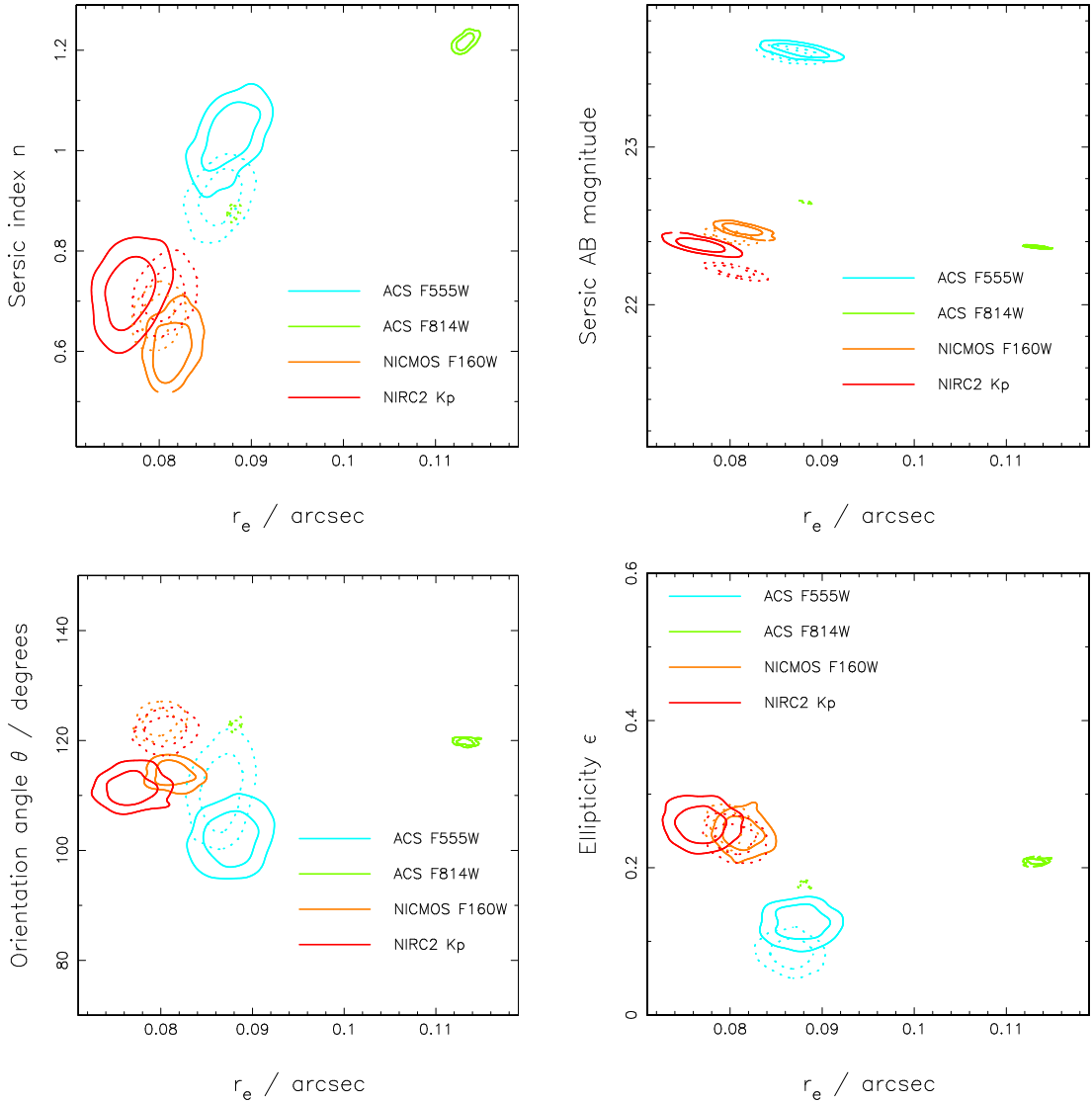
  
%
\begin{center}\begin{minipage}{0.9\linewidth}
  \plottwo{0737_b_re.vs.b_n.ps}{0737_b_re.vs.b_mag.ps}
\end{minipage}\end{center}
\begin{center}\begin{minipage}{0.9\linewidth}
  \plottwo{0737_b_re.vs.b_theta.ps}{0737_b_re.vs.b_eps.ps}
\end{minipage}\end{center}
\caption{Marginalized posterior probability distributions for pairs of 
source model parameters, given each dataset. 
Top left: effective radius \re and \sersic index $n$; 
Top right: effective radius \re and AB magnitude; 
Bottom left: effective radius \re and orientation angle $\phi$;
Bottom right: effective radius \re and ellipticity $\epsilon$.
%
The contours enclose 68\% and 95\% of the integrated probability.
Solid curves are for
the preferred galaxy subtraction scheme \subschemeone, while the dashed curves
are for the alternative scheme \subschemetwo.}
\label{fig:srcpars}
\end{figure*}

%
The top left-hand panel of Figure~\ref{fig:srcpars} shows the 2-d marginalized
probability distributions for two source model morphology parameters, the
effective radius \re and \sersic index~\sindex, given each of the datasets.  We
again note that the precision available for each parameter is much higher for
the deep \acs \Ifilter image, and very similar across the other three
datasets. Likewise, the lower panels in this figure show the inferred source
orientation and ellipticity, which are reasonably constant through the
bandpasses.

We infer a small, compact source galaxy across the whole wavelength range. 
%
The differences in morphology between the filters are not large, but there is a
suggestion that 
in the redder bands the profile is slightly
more compact, approaching the Gaussian distribution (\sindex$ \approx 0.7$).  
However, the degeneracy between \re
and~$n$ can be clearly seen, warning us not to over-interpret the inferences:
a robust conclusion is that the inferred \sersic index is low in all filters.
Likewise, the two different linestyle PDFs plotted also showing the effects of
the different lens galaxy subtraction schemes on the inferred source morphology.
In particular, the deep  \acs \Ifilter data can be seen, as expected, to be
generally more systematics-dominated than the other filters', with significant
(if small) differences in inferred effective radius and magnitude between
different analyses. It is in this filter that the sensitivity to the different
model assumptions is highest, and the limitations of our simply-parameterized
model are made most clear.  

The photometry is also (unsurprisingly) affected by the lens galaxy
subtraction: the lens subtraction systematic error can be seen in the
%
top right-hand panel of Figure~\ref{fig:srcpars}. In the next section we
use the photometry from subtraction scheme \subschemeone, and return
to the systematic error budget in section~\ref{sect:discuss}.


\subsection{Spectral energy distribution and stellar mass of the source galaxy}
\label{sect:SEDresults}

Armed with photometry from \acs (\Bfilter and \Ifilter), \nicmos
(\Hfilter) and \nirc (\Kfilter), we now reconstruct the spectral
energy distribution (SED) of the source. To account for uncertainty in
the zero points and filter transmission curves, we assert statistical
errors of 0.1 and 0.05 respectively and add these in quadrature to the
statistical errors from the MCMC inferences. 

\begin{figure}[!ht]  
\plotone{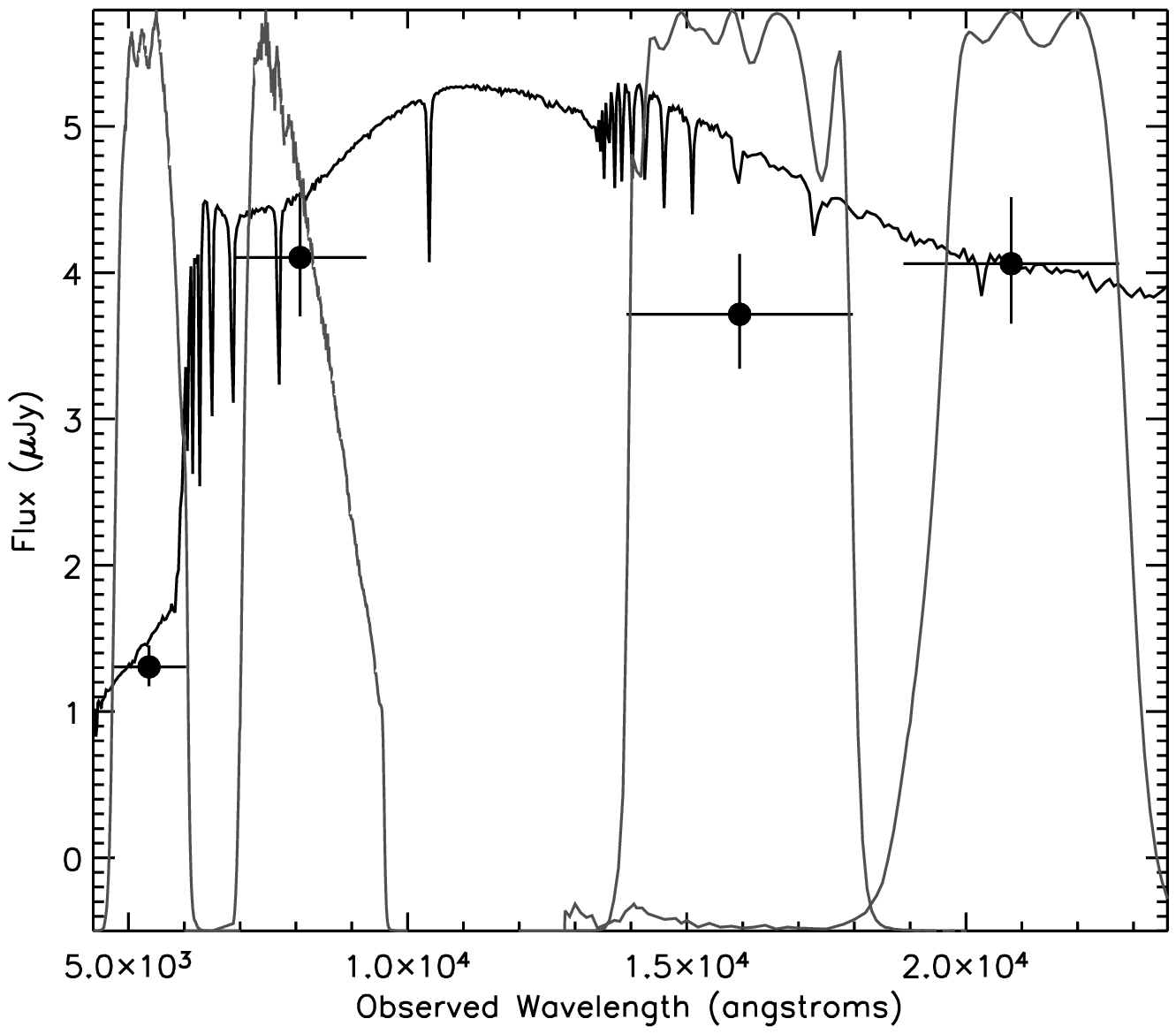}
\caption{Reconstructed source SED. 
%
The solid curve is the best-fitting spectrum normalized to the  
luminosity inferred in the \Kfilter filter; 
the error bars on the flux points show the statistical
errors assumed in the fit. In the background we show the filter
transmission curves (from blue to infra-red: 
\Bfilter, \Ifilter, \Hfilter, \Kfilter).}
\label{fig:sed}
\end{figure}

%
Given the known redshift, we estimate the stellar mass by fitting  
the observed colors to a variety of SED templates~\citep{Bun++06}.
The best fit model is obtained
for a exponential star-formation rate with short characteristic timescale
$\tau \sim 0.04$ Gyr and young age ($\lsim 0.7$ Gyr), and corresponds to a stellar mass
of \logstellarmass = 
(\logstellarmassvalue) assuming Chabrier IMF, where the error bar is
obtained by marginalizing the posterior over the stellar populations'
parameters. In other words, the system appears to have undergone a
very recent burst of star formation, consistent with its selection via
emission lines. 
This inferred star formation history 
is consistent with the SED-fitting performed by
\citet{Guz++03} on a sample of luminous compact blue galaxies taken from the
sample of \citet{Phi++97}. 

Figure~\ref{fig:sed} shows the 
fluxes (and uncertainties) used in the fit 
plotted as a function of wavelength, 
%
with the best-fitting galaxy template---normalized to the  
observed \Kfilter luminosity---overlaid.
For reference, the absolute AB magnitude in the \Bband is 
$-19.66\pm0.05$.
We note that choice of IMF is the single largest
source of systematic uncertainty \citep[0.2-0.3 dex, ][]{Bun++06} 
in the {\em absolute} stellar mass. However, when comparing stellar masses with
other surveys we must compute the same model-dependent masses. Both
\citet{She++03} and \citet{Bar++05} assume a Kroupa IMF, which results in
stellar masses different from those assuming a Chabrier IMF by just 0.05~dex.

Likewise, we note that the stellar masses of less well-resolved galaxies in the
literature typically also come from a global modeling of the object photometry
(rather than a joint morphological and photometric analysis), justifying our
approach to modelling the SED here. 
%
The \sersic indices measured in
section~\ref{sect:sourceresults} are also sufficiently similar to justify the
assumption of a single stellar population
when estimating the stellar mass. 
We do not, in any case, expect the systematic
error in the {\em absolute} stellar mass introduced to be greater than that from
the IMF uncertainty. Furthermore, the consistency between
the filters (all the way out to the \Kband) suggests that we are not dominated 
by small-scale star-forming regions in either the mass or size measurements.


\section{Systematic errors}
\label{sect:syserr}

Photometry with AO imaging has the reputation of being at best
difficult and at worst inaccurate. In this work we have looked
carefully at several systematic errors associated with photometric and
morphological of small extended objects viewed through galaxy-scale
gravitational lenses, and now discuss these errors in a little more
detail.


\subsection{Model-dependent LGSAO photometry}

The basic problem of measuring the total flux of an object, and the
radius within which half of this total flux is contained, is partially
solved by the assumption of a sensible model intensity distribution,
allowing the light profile to be extrapolated beyond the data
region. This solution is of course only as good as the model
assumption, but at least leads to a set of well-defined quantities
(\eg ``\sersic magnitudes''). The underlying assumption is that high
resolution imaging data provides enough constraints on the inner part
of the profile that the extrapolated quantities can be accurately
inferred. 

One could argue that imposing a model in this way ``biases'' the
results -- distant galaxies are not necessarily expected to have
pure \sersic profiles.  The system studied here at least appears to be simple,
in that a single image component provides a reasonable fit in the
infra-red, but there are suggestions in the bluer filters
that the galaxy has a more complex morphology. This is perhaps to be expected
given that this system was selected for its emission line spectrum, indicating
ongoing star-formation and consequent likely clumpy morphology.
However, if we are to
quantify galaxies like the source behind \lensname in a way that
permits comparison with other datasets and/or with a physical theory
then the \sersic profile appears to be the most natural choice, given
its widespread use.  The galaxy itself may not be well-fit by a \sersic
profile -- but that does not mean that knowing its \sersic parameters
is not useful. The fitting of a lensed \sersic profile is an appropriate
way of measuring the {\em average} properties of the source light
distribution, even in the blue filters. We note that the residual features in
the bluer images are smaller still than the inferred \sersic component,
suggesting that we are measuring the principal stellar structure, and 
not a smaller, brighter, peripheral star-forming region, even
at the shorter wavelengths.


\subsection{PSF model selection and truncation}

Assuming a model galaxy profile, and having 4 predictable copies of
the same image, means that the PSF structure can be inferred
concurrently with the source itself. Indeed, we have shown that PSF
selection via the Bayesian evidence is possible: there is information
in all the imaging data analyzed on the most appropriate PSF. We noted
that, since the number of model parameters is unchanged between the
different PSF models, the evidence is being dominated by the goodness
of fit. However, the PSF suitability is related to the choice of
source galaxy model and its parameter prior PDFs. This leads to the
evidence being a sharper tool for PSF selection than the reduced
chi-squared values, as can be seen in Table~\ref{tab:stats}.  In the
case studied here, the PSF selection is interesting but not critical,
as we clearly see that there are larger systematic effects at play.

Our treatment of the PSF with a small cutout star image is cause for
more concern. Our (internally-normalized) PSF postage stamps, at 16
pixels width, only span 1.5 times the seeing disk FWHM (\seeing). To
quantify the effect of this on the inferred model parameters, we
simulated \nirc observations of a gravitational lens having the same
properties as \lensname (\ie the parameter values found in
sections~\ref{sect:lensresults} and~\ref{sect:sourceresults}). For the
PSF we used a concentric sum of two Gaussians (representing the seeing
disk wings and the Airy pattern core, with relative weights given by
the Strehl ratio), following \citet{LSE06}.  The simulated data were
generated with a large 72-pixel PSF cutout, while the MCMC sampler was
provided with a posterior PDF assuming a small, renormalized 16-pixel
PSF cutout. Investigating input Strehl ratios of 0.2, 0.3, and 0.4
(and assuming FWHM values of 0.10 and 0.40 arcsec for the \Kband Keck
diffraction pattern core and seeing disk respectively), we found that
from a choice of model double Gaussian PSFs with the true seeing and
core size and Strehl ratios of 0.2, 0.3, and 0.4 the evidence selected
the correct (input) PSF each time, by about the same margin as seen
with the real data. Using the maximal evidence PSF, we then found that
the magnitude of the source was underestimated by 0.03 mags comparable
to the statistical uncertainty. This is significantly smaller than the
other estimated errors (that were used in the stellar mass
calculation), but comparable to the error introduced by the lens
galaxy subtraction. The effective radius was found to be
over-estimated by 0.005 arcsec, a small but statistically significant
increase; the \sersic index was also overestimated by 0.1 or so.  These
shifts, while contributing to the overall systematic error budget, do
not affect our conclusion about the unusual size of this source, which
we discuss below.


We conclude that LGSAO photometry of faint extra-galactic extended
sources at the 0.05 magnitude accuracy level (not including zero point
and filter curve calibration) is perfectly possible using techniques
such as those used in this work. However, we caution that the
conditions of observations were exceptionally good, both in terms of
seeing and stability of the PSF. This is supported by 
the fact that
the specially-observed star \psfzero gave the best results, and by the
consistency between the results obtained with this star and with the 
serendipitous stars
observed in the object field itself. This consistency is not guaranteed in
general, since the PSF can be expected to change significantly on timescales 
like 
the time interval between our observations of the lens field and of
\psfzero (J. Graham, priv.\ comm.), and that spatial variations of the
PSF can be significant (Steinbring et al, in prep.). 
However, it bodes well for the future that our results would be essentially
unchanged had we only used the stars in the field of the target.


\subsection{Overall systematic error budget}

In sections~\ref{sect:lensresults} and~\ref{sect:sourceresults} we identified
the lens galaxy subtraction as a serious issue leading to the dominant
systematic error when inferring the model parameters from well-calibrated data
and assuming an isothermal density profile lens.
A better approach would be to fit the lens and source  simultaneously, making
use of the quadruple-imaging to constrain the two intensity distributions with
minimal degeneracy.  From the Moffat profile fit residuals (scheme
\subschemetwo, Table~\ref{tab:stats}) we see that such a procedure would require
a  flexible model (such as the B-splines used here) for the lens galaxy light,
in order to get a good fit. This is  not computationally feasible within the
current framework, but should be possible in the semi-linear formalisms of
\citet{W+D03} and others.

Comparison of the parameter estimates between subtraction schemes does give a 
quantitative feel for the systematic errors introduced by the lens galaxy
subtraction. These are compared with the other errors identified in this work in
Table~\ref{tab:systematics}. 
We see that even the largest image analysis systematic error, that due to the
lens galaxy subtraction, is still smaller than that introduced by the assumption
of an isothermal density profile lens mass distribution.
Conservatively combining all the systematic errors by simple addition, 
the resultant systematic errors on the size and stellar mass
are approximately $0.1$ kpc and  $0.8 \times 10^9 \msun$; these may be
compared
with the statistical uncertainties shown in Figure~\ref{fig:sed}.

For the \sersic index we read off the systematic errors from
Figure~\ref{fig:srcpars} as~0.2 for the optical filters, and 0.1~for the
infra-red filters, and assume that this is unaffected by the lens density profile
(which simply changes the magnification of the source).

\begin{table}[!ht]
\caption{Summary of systematic errors identified in the text, and their effects
on the principal source parameters.}
\label{tab:systematics}
\begin{tabular}{cccc}
\hline
Description                     &  $\delta$\re & $\delta m_{\rm AB}$  & $\delta$\logstellarmass  \\
                                & (kpc)        &               & (dex)                    \\
\hline\hline
Incorrect stellar PSF model     & 0.01         & 0.01          & 0.005 \\
Truncated PSF image             & 0.005        & 0.03          & 0.015 \\
Lens galaxy subtraction         & 0.01         & 0.10          & 0.04 \\
Stellar mass IMF choice         &  N/A         &  N/A          & 0.05 \\
Lens model density slope        & 0.07         & 0.26          & 0.11 \\
Total (approximate)             & 0.10         & 0.40          & 0.17 \\
\hline
\end{tabular}
\end{table}


\section{Discussion: the size-mass relation at $z = 0.6$}
\label{sect:discuss}

\begin{figure*}[!ht]  
\plottwo{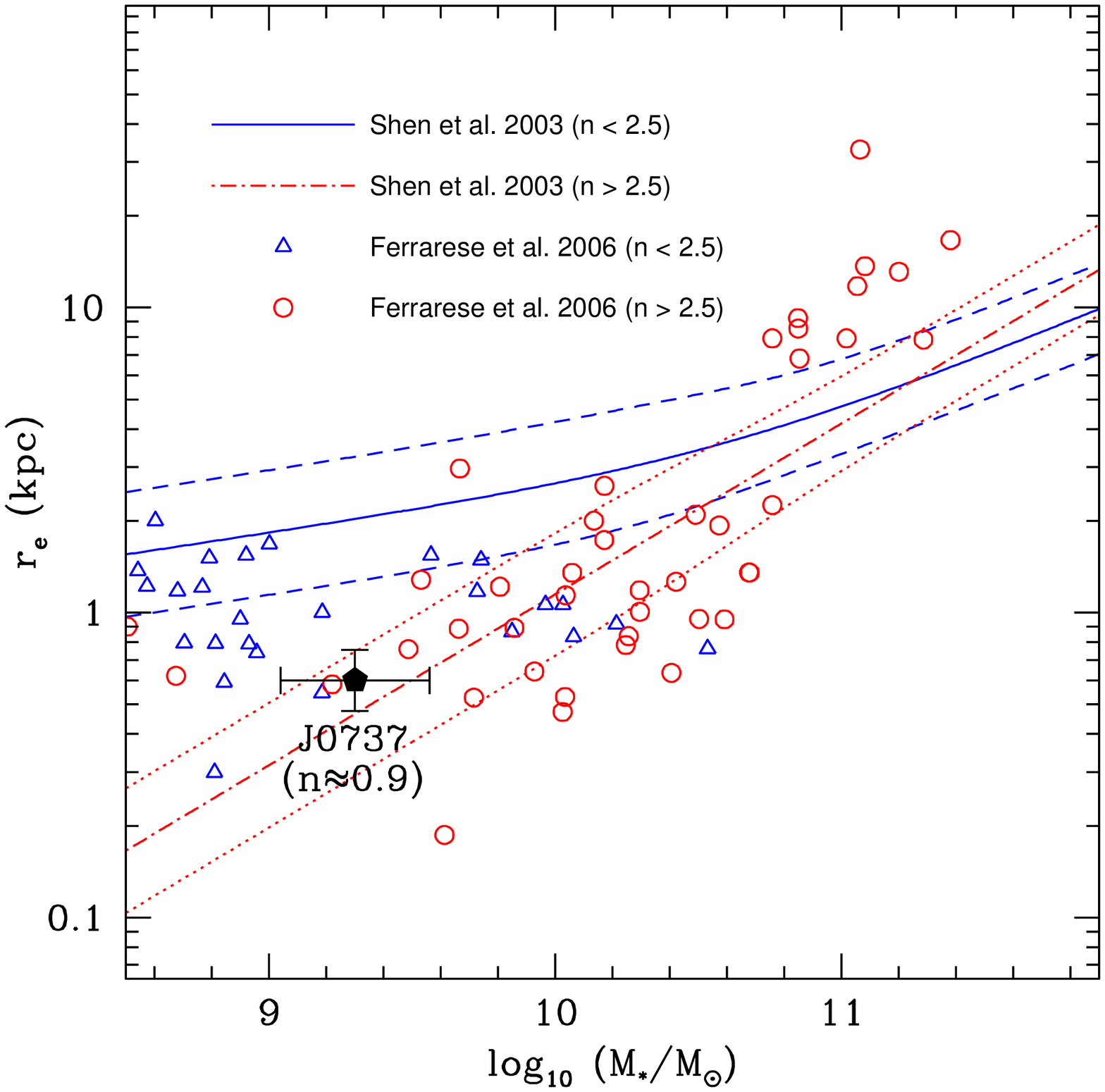}{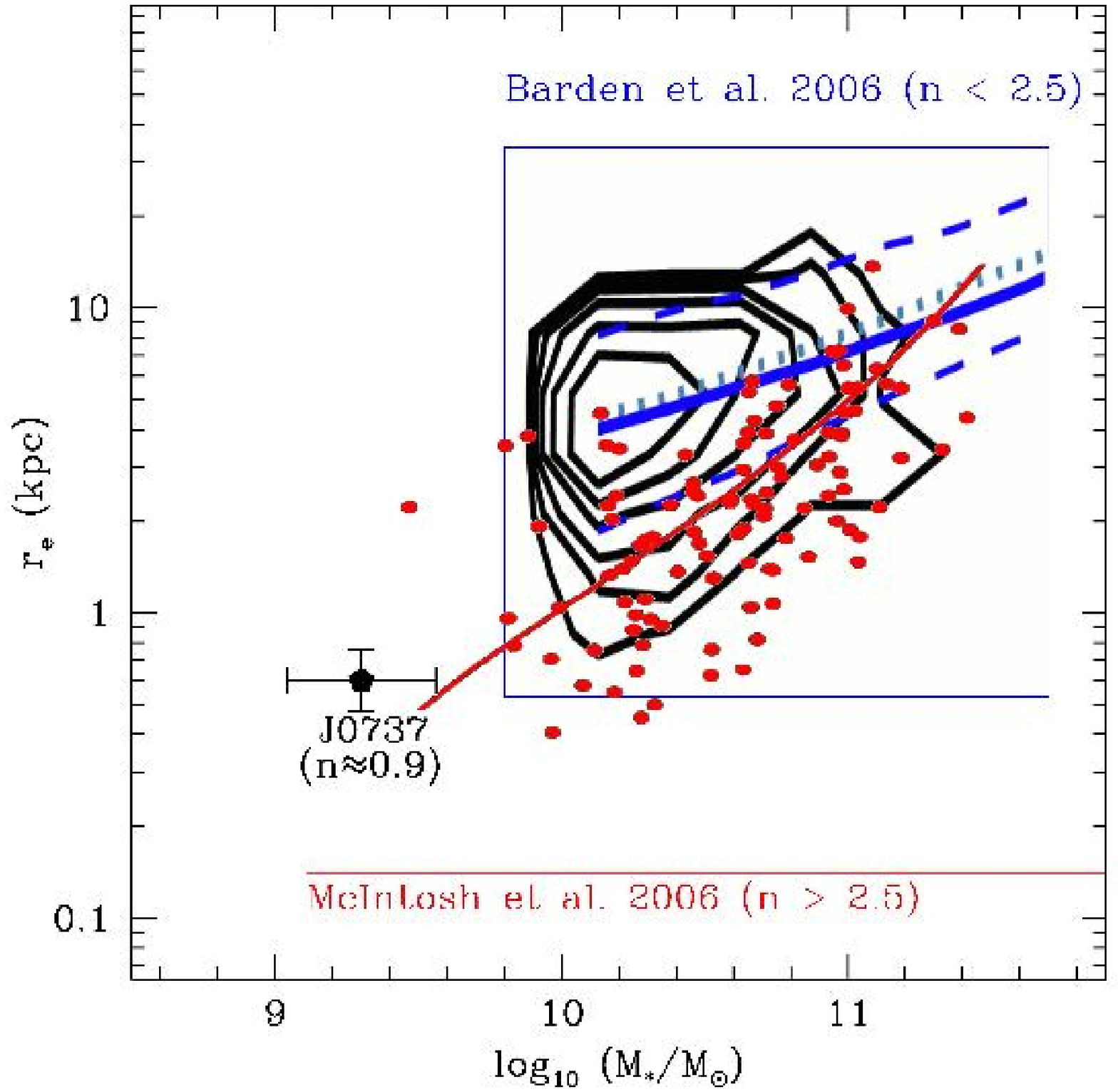}
\caption{Size-mass relations for galaxies selected by their measured \sersic
index. 
We plot the (relation between the) 
stellar mass and effective radius for galaxies 
in the local Universe \citep[left, SDSS, ][]{She++03} and at
$z\approx0.6$
\citep[right, GEMS, ][]{Bar++05,McI++05,Som++06}.
``Disk-like'' ($n<2.5$)
galaxies are shown in blue, 
and ``bulge-like'' ($n>2.5$) galaxies are shown in red.
In the left-hand plot we show the morpholgically-selected dwarf early-type
galaxies in the Virgo cluster, again divided by \sersic index. 
The black pentagonal point shows the source behind \lensname. 
The mass plotted here assumes a Kroupa IMF.}
\label{fig:smr}
\end{figure*}

From the analysis presented above we obtain the final result that \logre =
(\logrevalue), \logstellarmass = (\logstellarmassvalue) and \sindex =
(\bluesindexvalue) for the size, stellar mass, and \sersic index of the 
\lensname
%
source galaxy, where these numbers are global estimates based on all the
filters' data: the plots in Figure~\ref{fig:srcpars}, and the systematic error
analysis of the previous section indicate that the small differences between
the different bands are not significant.  
%
\footnote{We note that the SDSS size estimates were made in the r-band, such that the
effective rest-frame wavelength is about 5700~angstroms. At redshift 0.6, this
falls in the i-band, as used (approximately) in GEMS. 
Our corresponding \Iband measurement is the one most
affected by systematic errors -- however, 
the global size estimate we use can be seen
(Figure~\ref{fig:srcpars}) to be representative of the size in this filter.}
%
We overlay these values on the local relation for ``disk'' galaxies (\ie
those with \sersic index $n<2.5$) derived from SDSS by \citet{She++03}, and the
corresponding
$z\approx0.6$ relation derived from GEMS by \citet{Bar++05} \citep[and
interpreted by][]{Som++06} in Figure~\ref{fig:smr}. 
For its stellar mass (which
is a factor of 5 smaller than the GEMS completeness limit), the source behind
\lensname appears to be about a factor of $3$ smaller than the average local
galaxy \citep{She++03}, putting it approximately 3-$\sigma$ below the local
%
size-mass relation for galaxies with low \sersic index. 

The distant GEMS data do not extend to small enough masses to allow for a direct
comparison with our measurement. However, bringing our point into agreement with
the typical $z=0.6$ galaxy from GEMS would require a somewhat marked flattening
of the size-magnitude relation at masses lower than $10^{10} \msun$ since
redshift~0.6, which appears unlikely given the very modest evolution observed at
%
%
%
masses above $10^{10} \msun$: \citet{Bar++05} find a constant size-mass
relation;  \citet{Tru++06} measure \re$\sim(1+z)^{-0.4\pm0.06}$ for disk-like
galaxies, predicting that the mean object at $z\approx0.58$ is only about
$0.83$ times the size of the mean local disk-like galaxy. Even comparing with the
incomplete GEMS data at $\sim 10^{9} \msun$ \citep[][Figure 10]{Bar++05} we see that our
object is unusually small. Thus we conclude that the source galaxy behind
\lensname is, relative to existing surveys, somewhat extreme in terms of mass
and size, \emph{if it is indeed a disk galaxy}. 

Such compact galaxies have, however, been well-studied.  \citet{Koo++94}
identified a sample of compact, narrow emission-line galaxies in the
Hubble Deep field at redshift~$\approx 0.2$, having luminosities ($M_B
\approx -19$) and sizes (\re~$ \approx 1$~kpc). These objects comprise
a small fraction of the ubiquitous faint blue galaxies reviewed by
\citet{Ell97}.  Extending the sample to intermediate redshift and
focusing on the higher luminosity members ($M_B \approx -21$),
\citet{Koo++95} found using high resolution spectroscopy that these
objects appear more similar to local HII galaxies (dwarf galaxies
showing violent star formation activity), and suggested that these systems will
evolve into today's dwarf spheroids. This conclusion was also reached
by \citet{Phi++97} in an extension of that work, although they note
that the number densities are such that not all compact galaxies at
intermediate redshift can be progenitors of spheroids.  However,
\citet{Ham++01} argue that the observed narrow emission line widths
may not represent the depth of the whole galaxy potential, and instead
argue on the basis of the stellar masses they infer from their spectra
and photometry that the more luminous, compact galaxies are more
likely the progenitors of the bulges of present-day massive spiral
galaxies.  At the stellar mass scale inferred in this work
($10^{9.3}$), the source behind \lensname appears closer to the
low-luminosity end of the samples of \citet{Koo++95} and
\citet{Phi++97}, but is almost a factor of two smaller than the
limiting size of their sample ($0\farcs16$) and is fully resolved.
%
Indeed, our physical resolution is comparable to that reachable for galaxies in
the Virgo cluster with ground-based seeing; we find that the \lensname source
is comparable to the smallest dwarf ellipticals seen in the Virgo
cluster~\citep[][Figure~\ref{fig:smr}]{CCD92,Fer++06}, 
and is typical of the objects in the smallest size bin
of the \sersic-selected ``elliptical'' galaxies in 
the GEMS survey~\citep{McI++05}.
%

Can we interpret our super-resolved source morphology as being that of a forming
spheroid? The low \sersic index measured would suggest not, placing it firmly in
the disk-like samples of the literature. However, at low masses there is
evidence that elliptical galaxies can have \sersic indices of one and
below~\citep[e.g.\ ][]{TBB04}. The consistency in morphology between the
observation filters is indicative of a regular spheroid (although the residual
structure in the bluer filters would argue against a highly-evolved, smooth
stellar distribution. Perhaps the strongest indicator is the position of the
source in the size-mass plane. In Figure~\ref{fig:smr} we show the
local, and $z=0.6$, 
size-mass relations for ``elliptical'' ($n > 2.5$) galaxies from SDSS and
GEMS~\citep{She++03,McI++05}, and can see that the source behind \lensname sits
rather more comfortably with these relations, albeit at significantly (a factor
of 2 -- 4) lower mass.

Our results demonstrate that, using a gravitational telescope to
super-resolve the source, it is possible to study in considerable
detail atypical sources that may well be missed or excluded by
non-lensed surveys. In fact, size-mass studies at redshift~0.5 and
above have necessarily focused on the high luminosity end, and have
inevitably included a size cut to remove stars from the catalogs,
before a further completeness cut that discards the least massive
galaxies. It is not clear just how many small galaxies are being
overlooked in this way. The higher resolution afforded by our
gravitational telescope allows us to study the structure and surface
brightness profiles of the compact blue galaxies in much greater detail
and with higher precision, and to extend the investigation to smaller
sizes still.  For comparison, the total magnification provided by
\lensname is $\mu\approx13$, indicating an angular resolution in the
source plane of approximately 0.01~arcsec. The~$10\%$ accuracy we
obtain on our size measurement indicates that we are still some way
from the limit imposed by the resolution of our optics. Indeed we note
that this accuracy can be improved by a further factor of 2 by simply
using a more flexible lens model.

Having demonstrated the power of this method, a larger sample of
objects is needed in order to infer statistically meaningful
conclusions about the low mass/size tail of the mass-size relation.
As clearly discussed by \citet{Bar++05}, to achieve this goal it is
crucial to understand the selection function of the objects being used
in the size-mass relation study. Due to a form of the so-called
magnification bias, gravitational lens surveys such as SLACS tend to
favor compact sources. SLACS lenses were selected from (spectroscopic)
observations where the system is essentially unresolved, meaning that
lens systems with high total magnification are preferentially
detected. This bias is strongest when the source is
point-like, \ie much smaller than the size of the fiber and the
Einstein Radius ($\approx 1$~arcsec). Thus, it is not so surprising that the
first source to be studied is compact.  This realization implies that
when performing statistical analyses of the size-mass relation of
lensed galaxies it will be necessary to use Monte Carlo simulations to
understand and quantify the selection function of multiple image
systems. Applying our methodology to other lens systems, will, once
the selection effects are quantified, extend this study to join the
existing statistical analyses of higher-mass disks, to probe the small
size (\ie low angular momentum) regime.

The SLACS lenses are well-suited to this task: the efficiency of the
survey is such that some 100 high-magnification systems like \lensname
are expected to be found by the end of the program. (The number of
systems currently confirmed using high resolution imaging is already
close to this figure.)  Extending the study to sources at even higher
redshifts requires more lenses to be discovered at greater distances:
the SL2S survey \citep{Cab++07} is expected to discover $\sim100$
suitable systems, with sources at redshifts of~1.0 and
higher. However, the detection of these systems (via a ground-based
imaging survey in the $r'$- and $i'$-bands) will lead to different
selection effects than those present in the SLACS survey, and will
again require Monte Carlo simulations in order to understand them.

The different identification schemes of the SLACS and SL2S survey
introduce also a different selection effect in terms of stellar
population, which will have to be modeled and taken into account when
interpreting the results. SLACS sources are emission line selected and
will therefore be representative of actively star-forming galaxies,
directly comparable with galaxies selected in narrow band
surveys. SL2S sources are continuum-selected and contain a mix of
actively star-forming, post-starburst and quiescent stellar populations,
directly comparable with the galaxy population studied by wide field
HST surveys in similar broad bandpasses.

An additional implication of the lensing selection effect is that the
magnification bias in some sense increases the power of a galaxy
survey, by picking out the smallest objects and then making them
measurable.  With current technology, gravitational telescopes are the
only way of accurately measuring such tiny objects.


\section{Conclusions}
\label{sect:conclude}

We find that high quality images from \nirc are capable of providing very 
similar precision on simple lens and source model parameters to typical
datasets from \acs and \nicmos. The data themselves contain information about
the most appropriate PSF model to use, to the extent that a set of nearby
unsaturated stars can  be fruitfully compared using suitable statistics that
are  sensitive to the goodness-of-fit. We estimate that even for the LGSAO
imaging  this way of modeling the PSF allows a photometric precision of 0.05
mag.  
%
However, the calibration of isothermal 
However, the calibration of \emph{isothermal} 
galaxy-scale gravitational lenses as
cosmic telescopes is very likely limited by the subtraction of the lens
galaxy light. We estimate that this procedure introduces up to 0.1
magnitudes of systematic error into the source galaxy photometry. However, this
is still smaller than the error introduced by the assumption of an isothermal
density profile for the lens itself.

With this in mind we draw the following conclusions about the source behind
\lensname:

\begin{itemize}

%
\item Our photometry is robust enough to permit a reconstruction
of the SED, and we find a stellar mass of
(\stellarmassvalue). This is a factor of 5 smaller than the completeness
limit of the GEMS disk galaxy analysis of \citet{Bar++05}, and also smaller than
the least massive spheroid at this redshift studied by~\citep{McI++05}.

\item The \sersic profile parameters of the source can be measured to
high accuracy. We find an effective radius of (\revalue) ($\approx
0.09$ arcsec with $\sim10$\% accuracy), and a \sersic index of
(\bluesindexvalue) in the \Iband ($\sim$ rest-frame~B), and that these
values change little over the rest-frame optical range.

%
\item This very small galaxy lies approximately 3-sigma below the
local size-mass relation for disks. However, it shares the properties of the
smallest of the compact narrow emission line galaxies of
\citet{Koo++94}, and, despite its low \sersic index, 
is more typical of the dwarf early-type galaxies observed in the Virgo
cluster~\citep{Fer++06} and the
``elliptical'' galaxies studied 
by~\citet{McI++05} at high redshift. 

\end{itemize}
 
%
While the planned  statistical analysis of a large sample of lensed galaxies
will rely on the detailed understanding of the selection function, it is
clear that the magnifying effect of gravitational lenses allows us to extend
current size-mass studies to smaller sizes and
lower masses than would otherwise be available, posing fresh challenges to
models of galaxy formation and evolution.
%


\acknowledgments

We thank Laura Melling and Sherry Suyu for useful discussions when developing 
the lens modeling code, and are grateful to the anonymous referee for insightful
comments that led to some improvement of the paper.
PJM was given support by the TABASGO foundation in the
form of a research fellowship. 
TT acknowledges support from the NSF
through CAREER award NSF-0642621, and from the Sloan Foundation
through a Sloan Research Fellowship. 
The work of LAM was carried out
at Jet Propulsion Laboratory, California Institute of Technology under
a contract with NASA. 
LVEK is supported in part through an NWO-VIDI program subsidy (project \#
639.042.505). 
TT and PJM thank the Center for Adaptive Optics
for organizing the 2007 spring retreat, during which part of this work
was carried out.
This work was supported in part by the National Science Foundation
Science and Technology Center for Adaptive Optics,
managed by the University of California at Santa Cruz under
cooperative agreement AST 98-76783.
This research is supported by NASA through Hubble
Space Telescope programs SNAP-10174, GO-10494, SNAP-10587, GO-10798,
GO-10886, and in part by the National Science Foundation under Grant
No. PHY99-07949, and is based on observations made with the NASA/ESA
Hubble Space Telescope and obtained at the Space Telescope Science
Institute, which is operated by the Association of Universities for
Research in Astronomy, Inc., under NASA contract NAS 5-26555, and at
the W.M. Keck Observatory, which is operated as a scientific
partnership among the California Institute of Technology, the
University of California and the National Aeronautics and Space
Administration. The Observatory was made possible by the generous
financial support of the W.M. Keck Foundation. The authors wish to
recognize and acknowledge the very significant cultural role and
reverence that the summit of Mauna Kea has always had within the
indigenous Hawaiian community.  We are most fortunate to have the
opportunity to conduct observations from this mountain.


\bibliographystyle{apj}


\appendix
\section{The Effect of the Lens Mass Density Slope on the 
Inferred Source Size and Magnitude}

The local magnifying and distorting effect of a gravitational lens 
\citep[see\eg][]{Sch06} can be summarized by the (inverse) amplification matrix,
$\mathsf{A}^{-1}$:
\begin{equation}
\mathsf{A}^{-1} = \left( 
\begin{array}{ccc}
1 - \kappa + \gamma & 0 \\
0                   & 1 - \kappa - \gamma \end{array} 
\right),
\label{eq:amatrix}
\end{equation}
where $\kappa$ and $\gamma$ are two combinations of the spatial 
second derivatives of the projected gravitational potential -- $\kappa$ is
proportional to the projected (surface) mass density -- in a Cartesian coordinate system aligned with the radial and
tangential directions. To first order, a 
source of width $dx$ and length $dy$ (also aligned with these axes) 
is distorted into an image of width $dx_i$ and $dy_i$ according to
\begin{equation}
\mathsf{A} \left(\begin{array}{c} dx \\ dy \end{array}\right) = 
\left(\begin{array}{c} dx_i \\ dy_i \end{array}\right).
\end{equation}
The $\mathsf{A}_{11}$ component describes the radial stretching of the
source, while the $\mathsf{A}_{22}$ component describes the tangential 
stretching. The factor by which the solid angle subtended is increased 
due to the lensing
effect is the magnification $\mu = |\mathsf{A}| = 1/|\mathsf{A}^{-1}|$.

In terms of the Einstein
radius ($\theta_{\rm E}$, the radius at which the 
magnification is formally infinite), the 
quantities $\kappa$ and $\gamma$ are
given by 
\begin{equation}
\kappa = \frac{2-m}{2} \left( \frac{\theta_{\rm E}}{\theta}\right)^m ;\;\;\;\;\;\;\;
\gamma = \frac{m}{2} \left( \frac{\theta_{\rm E}}{\theta}\right)^m. 
\label{eq:kappagamma}
\end{equation}
%
%
for a simple spherically-symmetric lens with power-law density profile. 
for a simple, spherically-symmetric, power-law density profile lens (with
logarithmic slope~$m$). 
Two images form at positions $\theta_{\pm}$ that solve the lens equation, 
\begin{equation}
\beta = \theta_{\pm} - \alpha(\theta_{\pm}), \;\; \text{where, in this case,}\;\; \alpha(\theta_{\pm}) = \left[ \kappa(\theta_{\pm}) + \gamma(\theta_{\pm})\right] \theta_{\pm}.
\end{equation}
If the source position~$\beta << \theta_{\pm}$, 
as is the case when the images are highly magnified and
are close to forming an Einstein ring, we find the images at 
\begin{equation}
\theta_{\pm} \approx \theta_{\rm E} ( 1 \pm \epsilon ) \;\; 
\text{where} \;\; \epsilon = \frac{\beta}{m\theta_{\rm E}}.
\label{eq:solves}
\end{equation}
The offset~$\epsilon$ is well-constrained by the data, and so we
proceed treating $\epsilon$ as a small ($<<1$) constant.
At this point we note that the image positions and distortions do contain
some information
on the density slope~$m$, allowing this parameter to be fitted. What we are
working towards here is a quantification of the effect of perturbing the 
slope~$m$ away from the isothermal value ($m=1$).

Evaluating $\kappa$ and $\gamma$ at the image postions, substituting into
equation~\ref{eq:amatrix} and expanding to first order in~$\epsilon$ we find
that
\begin{equation}
\mathsf{A}^{-1} \approx \left( 
\begin{array}{ccc}
m ( 1 \pm \epsilon \mp m \epsilon ) & 0 \\
0                                   & \pm m \epsilon \end{array} 
\right),
\label{eq:amatrix2}
\end{equation}
and that the inverse magnification is (also to first order) 
$\mu^{-1} \approx \pm m^2 \epsilon$. 

We can now use this result to estimate the uncertainty on the inferred source
size (denoted by $\sigma_{r_{\rm e}}$) 
given by a systematic error in the model slope~$m$. We first note that the
inferred source plane solid angle is given by 
\begin{equation}
\Omega = dx \cdot dy = \Omega_{\pm}  \mu^{-1}(m), 
\end{equation}
where $\Omega_{\pm}$ is the solid angle subtended by each image, and $\Omega
\sim r_{\rm e}^2$. A small change in the density
slope away from a fiducial value of 1 gives rise to an error in 
source area $\Omega$ according to
\begin{eqnarray}
\sigma_{\Omega} &= \Omega_{\pm} \left. \frac{\partial \mu^{-1}}{\partial m} \right|_{m=1} \sigma_{m},\\
\text{such that}\;\; \frac{\sigma_{\Omega}}{\Omega} &= \frac{1}{\mu^{-1}} \left. \frac{\partial \mu^{-1}}{\partial m} \right|_{m=1} \sigma_{m}.  
\end{eqnarray}
From this, and the result above, we get that
\begin{eqnarray}
\frac{\sigma_{\Omega}}{\Omega} &\approx 2 \sigma_{m}, \\
\text{and so}\;\; \frac{\sigma_{r_{\rm e}}}{r_{\rm e}} &\approx \sigma_{m}.   
\end{eqnarray}
Since gravitational lensing conserves surface brightness, the inferred source
flux is simply proportional to the inferred source solid angle~$\Omega$:
consequently, the error in the AB magnitude due to uncertainty in the density
profile slope is 
$\sigma_{m_{\rm AB}} = (2.5/\log_{\rm e} 10)(\sigma_{\Omega}/\Omega) \approx 2.2 \sigma_{m}$.

\citet{Koo++06} give $\sigma_m = 0.12$ for the intrinsic spread of power-law
indices, where the profile is constrained at two radii, the Einstein radius and
the (smaller) effective radius. 
%
(Note that this small scatter was not appreciated by e.g.\ \citet{KRG01} in
their analysis of magnification errors.)
While the power law index they quote 
is not quite the local slope at the Einstein radius that we require here, the
range of radii they consider brackets the Einstein radius of \lensname and
therefore their value for $\sigma_m$ 
provides an approximate quantification of the size of the density slope 
systematic error.


\end{document}